\documentclass[12pt]{article}

\usepackage{amssymb,amsmath,amsthm,thmtools,amsfonts,bbm,eurosym,geometry,ulem,graphicx,caption,color,setspace,comment,footmisc,caption,natbib,pdflscape,array,hyperref,enumerate,nomencl,etoolbox,mathtools,booktabs,makecell,subcaption,algorithm}
\usepackage{algpseudocode}

\mathtoolsset{showonlyrefs=true}
\theoremstyle{remark}
\newtheorem{remark}{Remark}

\newcommand{\RR}{\mathbb{R}}
\newcommand{\ONE}{\mathbbm{1}}
\newcommand{\cC}{\mathcal{C}}
\newcommand{\cO}{\mathcal{O}}
\newcommand{\cQ}{\mathcal{Q}}
\newcommand{\bS}{\mathbb{S}}
\newcommand{\barbS}{\bar{\mathbb{S}}}
\newcommand{\bbE}{\mathbb{E}}
\let\oldparagraph=\paragraph
\renewcommand\paragraph[1]{\oldparagraph{#1.}}
\renewcommand{\today}{\ifcase \month \or January \or February \or March \or %
April \or May \or June \or July \or August \or September \or October \or November \or %
December \fi \number \year}

\title{DeepHAM: A Global Solution Method for Heterogeneous Agent Models with Aggregate Shocks}
\author{Jiequn Han\thanks{Flatiron Institute and Princeton University. Email: \url{jiequnhan@gmail.com}.},~ 
Yucheng Yang\thanks{Princeton University. Email: \url{yuchengy@princeton.edu}.},~
and Weinan E\thanks{Peking University and Princeton University.}
}

\date{This version: \today\\
~~First version: December 2021
}

\begin{document}
\maketitle

\begin{abstract}
    An efficient, reliable, and interpretable global solution method, the \textit{Deep learning-based algorithm for Heterogeneous Agent Models (DeepHAM)}, is proposed for solving high dimensional heterogeneous agent models with aggregate shocks.
    The state distribution is approximately represented by a  set of optimal generalized moments. Deep neural networks are used to approximate the value and policy functions, and the objective is optimized over directly simulated paths. In addition to being an accurate global solver, this method has three additional features. First, it is computationally efficient in solving complex heterogeneous agent models, and it does not suffer from the curse of dimensionality. Second, it provides a general and interpretable representation of the distribution over individual states, which is crucial in addressing the classical question of whether and how heterogeneity matters in macroeconomics. Third, it solves the constrained efficiency problem as easily as it solves the competitive equilibrium, which opens up new possibilities for studying optimal monetary and fiscal policies in heterogeneous agent models with aggregate shocks.
\end{abstract}


{\let\thefootnote\relax\footnotetext{\hspace*{-0.3em}We are grateful to Gianluca Violante for numerous discussions and constructive feedback throughout this project. We thank Mark Aguiar, Anmol Bhandari, Mahdi Kahou, Nobu Kiyotaki, Moritz Lenel, Ziang Li, Ben Moll, Jonathan Payne, Jesse Perla, Mikkel Plagborg-Møller, Simon Scheidegger, Chris Sims, Jeffrey Sun, Wei Xiong, and many seminar participants for helpful comments. }}

\section{Introduction}
The incorporation of both explicit heterogeneity and aggregate fluctuations into quantitative models has been one of the most important recent developments in macroeconomics. It has become an important agenda for a number of reasons. First, uneven dynamics across sectors and population groups after major economic fluctuations and policy shocks suggest that heterogeneity and aggregate shocks are necessary considerations when studying fundamental macroeconomic problems. Second, the recent development of the heterogeneous agent New Keynesian (HANK) models suggests that heterogeneity gives rise to key channels in the transmission of aggregate shocks. These channels are crucial in determining the correct aggregate implications of monetary and fiscal policies. Third, the advancement of computational methods and growth in computing power have allowed economists to build models more realistic than the representative agent models that have long been dominant in both academic and policy research.

Despite the attention this agenda has received, its workhorse models, heterogeneous agent (HA) models with aggregate shocks, still present severe computational challenges. Ideally, one would like to develop solution methods that fulfill the following basic requirements:
\begin{itemize}
\item \textbf{Efficiency}: The method should be computationally efficient, especially for complex HA models with multiple state variables.
This is necessary in order to use the method for calibration, estimation, and further quantitative analysis.

\item \textbf{Reliability}: The method should produce accurate solutions for all practical situations that HA models are intended for.
In particular, it should be applicable beyond the local perturbation regime if nonlinear and nonlocal effects of aggregate shocks are important.

\item \textbf{Interpretability}: We are not only interested in the numbers that come out of an algorithm, but also in understanding the mechanisms underlying the results.  For that, the major components of the algorithm should be interpretable.  In particular, solutions to HA models with aggregate shocks usually involve mappings from the distribution over all individual states to the agent's welfare or decision outcomes. An ideal solution should provide interpretability of these mappings through an interpretable representation of the state distribution. 
An interpretable representation of the distribution is also necessary to derive reduced dynamic models at the aggregate level from the original HA model.

\item \textbf{Generality}: The method should in principle be applicable to a wide variety of different HA models (simple or complex), and to different notions of equilibrium (e.g. competitive equilibrium or constrained efficiency problem). 
\end{itemize}

Currently, there are two main approaches for solving HA models with aggregate shocks, and they satisfy only a subset of the requirements
listed above. The first is the Krusell-Smith (KS) method, a global solution method proposed in \cite{krusell1998income}. The KS method approximates agent distribution with a small number of moments (e.g., the first moment), which are interpretable. It is efficient when solving simple HA models but becomes less effective when solving complex HA models with multiple shocks, or multiple endogenous states, or when solving the estimation problem. This is due to the large number of variables (such as the possibly large number of moments needed) introduced and the resulting curse of dimensionality problem. That is, the computational cost increases exponentially with the number of variables.

The second approach is the local perturbation method proposed in \cite{reiter2009solving}. This method allows one to study or estimate complex HA models, but is not reliable for models where aggregate shocks bring significant nonlinear or nonlocal effects.
Nonlinear effects are common in models with zero lower bounds (ZLB). Nonlocal effects appear in models with large aggregate shocks, or in models (e.g., macro-finance models) where explicit consideration of aggregate uncertainty plays an important role in shaping agents' behavior, resulting in the deviation of the risky steady state from the deterministic steady state. 
Table \ref{table:method_principle} summarizes the advantages and limitations of these two methods.

\begin{table}[ht!]
\centering
\begin{tabular}{lccc}
\hline\hline
Model features           & KS method & Perturbation method &  DeepHAM \\ \hline
Multiple shocks        & No                   & Yes   & Yes                 \\ 
Multiple endogenous states & No                   & Yes   & Yes         \\
Large shocks           & Yes                  & No       & Yes      \\
Risky steady state    & Yes                  & No       & Yes       \\ %
Nonlinearity (e.g., ZLB)  & Yes                  & No        & Yes    \\ \hline\hline
\end{tabular}
\caption{Model features that different solution methods for HA models with aggregate shocks can handle.}
\label{table:method_principle}
\end{table}

In this paper, we propose a new solution method, the {\it Deep learning-based algorithm for Heterogeneous Agent Models (DeepHAM)}, which satisfies all the requirements listed above. 
We formulate a HA model with $N$ agents, where $N$ is large when we aim to solve a problem with a continuum of agents.
To solve HA models, the fundamental objects of interest are agents' value and policy functions.
A complication arises from the fact that these functions depend not only on the agent's own state, but also the distribution of all agents' states in the economy. To address this issue, we represent the value function and policy function with deep neural networks, and present an algorithm to update the value and policy functions iteratively. Deep neural networks are a class of functions in deep learning, which have a strong representational capability for high dimensional functions and can be efficiently optimized with stochastic gradient descent algorithms.
In contrast to existing literature that uses deep learning to represent high dimensional policy and value functions directly \citep{maliar2021deep,azinovic2022deep}, we introduce \emph{generalized moments} to represent the state distribution efficiently, and solve for the value and policy functions as functions of the generalized moments. Generalized moments extract useful information from the state distribution, similarly to classical moments, but are represented by neural networks and automatically determined by the algorithm. The introduction of generalized moments also ensures that the agent's optimal policy and value functions are invariant under permutations of the ordering of the agents. Conceptually, the generalized moments reduce the state dimension while remaining readily interpretable and flexible enough to encode the state distribution through algorithmically-determined moments. In addition, the generalized moments share a similar representation to quantities such as the first moment of wealth distribution that are typically observed in the interaction between the agents and the whole economy.
As we will see below, a single generalized moment not only leads to more accurate solutions than using only the first moment, but also extracts key information from the agent distribution with first order implications for aggregate welfare and dynamics. Thus, it provides a general and interpretable way to study a key question in macroeconomics: whether, why, and how inequality matters for the macroeconomy.\footnote{Compared to the classical polynomial approximation used in the literature \citep{aruoba2006comparing,fernandez2016solution}, the neural network serves the same purpose as a function approximator. The difference is that the neural network does not rely on a fixed set of basis functions and can approximate high dimensional functions more efficiently.} 

As we will demonstrate later, DeepHAM  meets all the requirements listed above.
First, it shows better global accuracy compared with existing methods. In the baseline model we study, DeepHAM with only the first moment in the state vector reduces the Bellman equation error by 37.5\% compared to the KS method. DeepHAM with one generalized moment reduces the error by 54.2\%. 
Second, the computational cost of DeepHAM is quite low in solving complex HA models, and it does not suffer from the curse of dimensionality. DeepHAM can efficiently solve HA models with endogenous labor supply, or with a \cite{brunnermeier2014macroeconomic} type of financial sector. Third, the use of generalized moments allows us to revisit classical questions in macroeconomics of whether and how heterogeneity matters to aggregate welfare and dynamics. \cite{krusell1998income} famously argued that, in their setup, individual welfare is affected by other agents only through the mean of wealth distribution. With the generalized moments, we find that an unanticipated redistributional policy shock would have a non-zero welfare impact on those households who are not in the policy program, even when the mean of the wealth distribution is not affected. Finally, as a demonstration of the generality of DeepHAM, we show that it can be used to solve the constrained efficiency problem in HA models, which is regarded as a challenging problem in the literature, as easily as solving the competitive equilibrium. This allows us to study optimal fiscal and monetary policy in HA models with aggregate shocks.

DeepHAM should be applicable to a large class of economic problems with heterogeneity and aggregate shocks, and here we list some such examples. First, since DeepHAM does not suffer from the curse of dimensionality with more endogenous states or shocks, we can introduce more realistic portfolio options like housing and mortgage choices \citep*{kaplan2020housing, boar2021liquidity}. We can efficiently handle models with \textit{ex ante} heterogeneous agents, such as households and financial experts \citep{brunnermeier2014macroeconomic}, rational and bounded-rational agents \citep{woodford2021fiscal}, among others. We can study models with rich firm heterogeneity and aggregate shocks \citep{khan2013credit}. We can also study HA models with multiple shocks, where the shocks take forms that appear commonly in the DSGE literature \citep{mckay2016role}. Second, we can use DeepHAM to study models with large shocks such as the COVID-19 shock, or large endogenous fluctuations such as those discussed in \cite*{petrosky2018endogenous}. We can also use DeepHAM to study asset pricing and the wealth effects of monetary and fiscal policy in a HA model. The empirical literature has shown these factors to be important \citep{andersen2021monetary} but, due to computational challenges, they have only been studied in models with limited heterogeneity \citep{kekre2020monetary, caramp2021monetary}. We can also study the interaction of asset pricing and wealth inequality \citep{cioffi2021heterogeneous}. Third, we can study optimal policy problems with heterogeneous agents using the Ramsey approach \citep{davila2012constrained,nuno2018social}, such as optimal monetary and fiscal policy \citep*{bhandari2021inequality,dyrda2021optimal,le2021should}, or optimal macroprudential policy \citep{bianchi2018optimal}. Such study has heretofore been limited by computational challenges. Last but not least, methodologically, we can extend  DeepHAM to do model calibration by introducing a calibration target in the objective function so that we can solve and calibrate HA models in the same algorithmic framework.

\paragraph{Related Literature} Our work builds on an extensive literature on solving HA models with aggregate shocks. As discussed, there are two main approaches in the literature \citep{algan2014solving}: the global Krusell-Smith (KS) method \citep{krusell1998income, den2010comparison, fernandez2019financial, schaab2020micro}, and the local perturbation method \citep{reiter2009solving, winberry2018method, ahn2018inequality, boppart2018exploiting, bayer2020solving, auclert2021using}. Due to the curse of dimensionality, the KS method cannot handle complex HA models with multiple assets and multiple shocks. The perturbation method has been applied to complex HA models, for solving local dynamics around the deterministic stationary equilibrium in the absence of aggregate shocks, or for parameter estimation \citep{liu2019full}.
However, the perturbation method is inapplicable to problems with nonlinear dynamics induced by aggregate shocks, or problems that are not close to the deterministic stationary equilibrium. DeepHAM can handle complex HA models with aggregate shocks and provide a global solution.

This paper proposes a general methodology that extracts the key information of the distribution that matters for aggregate welfare and dynamics. This is an important issue in studying the role of heterogeneity in macroeconomics \citep{kaplan2018monetary, auclert2019monetary}. For an overview of this literature, see \cite{kaplan2018microeconomic}. Most papers in this literature study the role of heterogeneity with quantitative decomposition after solving the model \citep{kaplan2018monetary}, or with sufficient statistics that are derived analytically based on the first-order approximation \citep{auclert2019monetary}. In contrast, we propose a general numerical method that extracts key {generalized moments} of the distribution as part of the numerical solution process. These {generalized moments} can be viewed as a set of ``numerically determined sufficient statistics'' of the model. Our idea of the permutation invariant {generalized moments} coincides with the independent and contemporaneous work of \cite{kahou2021exploiting}, while we further explore the interpretation of the generalized moments and heterogeneity, and use them to study the impact of an unanticipated redistributional policy shock.

This work is also of relevance to the literature on machine learning-based algorithms for solving high dimensional dynamic programming problems in scientific computing \citep{han2016deep, han2018solving, fernandez2020solving} and in macroeconomics \citep*{duarte2018machine, fernandez2019financial, scheidegger2019machine, maliar2021deep, maliar2022deep, azinovic2022deep}. Recent notable contributions by \cite{maliar2021deep, azinovic2022deep} also use deep learning to solve HA models with aggregate shocks. DeepHAM differs from their work in the following aspects. First, we introduce generalized moments to make the high dimensional policy and value functions permutation invariant to the ordering of agents, which also improves interpretability. Second, in the recursive formulation, DeepHAM addresses the optimization problem with directly simulated paths, while \cite{maliar2021deep} optimizes over an objective function constructed as a weighted sum of the Bellman residual and the first-order condition. Their setup requires a good approximation not only of the high dimensional value function itself, but also of partial derivatives of the value function, which are challenging to accurately obtain using neural networks. \cite{azinovic2022deep} formulates the objective function as the weighted sum of the deviations from equilibrium conditions, which differs from our approach as well.
In addition, this paper presents the first example of the use of machine learning-based algorithms to solve constrained efficiency problems in HA models with aggregate shocks.

The rest of this paper is organized as follows. Section \ref{sec:general_setup} presents the DeepHAM method for solving a general HA model with aggregate shocks. Section \ref{sec:KS} illustrates the use of DeepHAM on the classic Krusell-Smith model, and highlights the main features of the current approach. Sections \ref{sec:jfv} and \ref{sec:planner} apply DeepHAM to more complex HA models and the constrained efficiency problem in HA models with aggregate shocks. Section \ref{sec:conclusion} concludes the paper with some perspectives.

\section{DeepHAM: A New Solution Method}\label{sec:general_setup}
We first present the DeepHAM method to solve the competitive equilibrium in a general HA model in Sections \ref{sec:game_setup} to \ref{sec:general_algo}. Then we extend our setup to the constrained efficiency problem in Sections \ref{sec:general_constrained} and present the DeepHAM algorithm accordingly.

\subsection{General Setup of HA Models}\label{sec:game_setup}
Consider a discrete time and infinite horizon economy consisting of $N$ agents.
$N$ is large when we aim to solve a problem with a continuum of agents, and smaller when we aim to solve for the strategic equilibrium with finite agents.\footnote{Although our model assumes a finite number of agents, numerical results suggest that a choice of $N=50$ can approximate the solution to HA models with a continuum of \textit{ex ante} homogeneous agents as in \cite{krusell1998income} quite well. We have also tested values of $N = 100, 200, ...$, with similar results. In the more general case, a proper choice of $N$ may depend on the nature of the model. }

For agent $i$, her state dynamics (i.e., law of motion for individual states) which usually come from the agent's budget constraint, are given by:
\begin{equation}
\label{eq:state_general_dynamics}
    s^i_{t+1} = f(s^i_{t}, c^i_{t}, X_t, \barbS_t; z^i_{t+1}), \quad i=1,\dots,N.
\end{equation}
Here $s^i_t\in\RR^{d_s}$ and $c^i_t\in\RR^{d_c}$ denote the state and decision (control) of agent $i$ at period $t$. $z^i_{t}\in\RR^{d_z}$ denotes the idiosyncratic shock at period $t$, and is a subvector of $s^i_t$. The set 
$
   \barbS_t = \{(s^1_t,c^1_t), (s^2_t,c^2_t) \dots, (s^N_t,c^N_t)\}
$
denotes the (unordered) set of all agents' state-control pairs. 
Similarly, we use $\bS_t=\{s^1_t,s^2_t,\dots,s^N_t\}$ to denote the set of agent states.
We will also frequently use $S_t=(s^1_t, s^2_t, \dots, s^N_t)$ to denote the (ordered) vector of agent states. $X_t\in\RR^{d_X}$ denotes aggregate state variables excluding $\bS_t$. Here, the law of motion of agent states \eqref{eq:state_general_dynamics} combines the agent's budget constraint, together with other optimization and market clearing conditions that characterize the aggregate prices in the budget constraint.
For example, in \cite{krusell1998income}, the household's wealth state in the next period depends on current wealth and consumption, as well as current prices of labor and capital.
$\barbS_t$ is included as an input to the law of motion $f$ for the individual state, because prices are functions of (the mean of) the wealth distribution according to the representative firm's optimization conditions and market clearing conditions, and the wealth distribution is a subset of $\barbS_t$.

The control variables are subject to inequality constraints,
\begin{equation}
\label{eq:general_borrowing_constraint}
    h_l(s^i_t, X_t, \bS_t) \leq  c^{i}_{t} \leq h_u(s^i_t, X_t, \bS_t), \quad i=1,\dots,N,
\end{equation}
where $h_l,h_u$ are vector functions with $d^c$-dimensional output, and the dynamics of $X_t$ are modeled by
\begin{equation}
\label{eq:exo_general_dynamics} 
X_{t+1}=g(X_t, \barbS_t;Z_{t+1}),
\end{equation}
where $Z_t\in\RR^{d_Z}$ denotes the aggregate shock, and is usually a subvector of $X_t$. The aggregate state variable $X_t$ also includes other aggregate quantities that affect the law of motion for individual states \eqref{eq:state_general_dynamics}, but cannot be written only as functions of the collection of state-control pairs $\barbS_t$. An example of $X_t$ that contains more information than $Z_t$ is presented in Section \ref{sec:jfv}.

Here we have indicated that $f,g, h_l,h_u$ depend only on the sets $\barbS_t$ or $\bS_t$, not the ordering of the agents,
i.e., the state dynamics \eqref{eq:state_general_dynamics} are invariant to the ordering of the agents in the economy.
This is a consequence of the ``mean-field'' character of the interaction between agents.
“Mean-field” is a concept that originated in physics, and describes the situation when the agents, or particles, interact with each other not directly, but through an empirical distribution that all the agents contribute to equally.
A typical interaction form of mean-field type is through endogenous aggregate variables $O_t\in\RR^{d_O}$ defined by
\begin{equation*}
    O_t = \frac1N \sum_{i=1}^N \cO(s^i_t, c^i_t).
\end{equation*}
Here $O_t$ can be considered a function of $\barbS_t$ as it is permutation invariant to the ordering of agents. Examples of $O_t$ include the first or other moments of individual states. 
In this paper, we assume the dependence of $f, g, h_l, h_u$ on $\barbS_t$ or $\bS_t$ can be written in explicit functional forms. The Krusell-Smith model mentioned above is such an example.
This point will be more clear in our presentation of the concrete examples in Sections \ref{sec:KS} to \ref{sec:planner}.\footnote{The assumption that we have explicit function forms of $f, g, h_l, h_u$ excludes HA models with nontrivial market clearing condition \citep[Section 4]{algan2014solving}. We leave it for further investigation in a companion paper.}

According to the above description, $(X_t, \bS_t)$ completely characterizes the state of the whole economy. Mathematically, we are interested in how agents should make decisions $c^i_t= \mathcal{C}(s^i_t, X_t, \bS_t)$ through the decision rule $\mathcal{C}$ to achieve optimality.\footnote{Note here $\mathcal{C}$ is common across agents. This naturally applies for HA models with \textit{ex ante} homogeneous agents like \cite{krusell1998income}. Similarly, the value function $V$ defined later is common across agents as well. For models with \textit{ex ante} heterogeneous agents, we may introduce additional individual state variables such that $\mathcal{C}$ is common across agents.}

In the setting of competitive equilibrium, each agent $i$ seeks to
maximize her discounted lifetime utility:
\begin{equation}
    \bbE_{\mu} \sum_{t=0}^{\infty}\beta^tu(c^i_t).
\end{equation}
Here $u(\cdot)$ is the utility function and $\beta \in (0, 1)$ is the discount factor. $\mu$ is the full state distribution $(X_0, \bS_0)$ at the initial time. We use $\mu(\mathcal{C})$ to denote the stationary distribution of $(X_t, \bS_t)$ when every agent employs the decision rule $\mathcal{C}$ and 
we assume that such a stationary distribution always exists. The expectation is taken with respect to both the idiosyncratic and aggregate shocks over all time.

We say that $\mathcal{C}^*$ is an optimal policy in the competitive equilibrium if, for $\forall i \in \{1, \dots, N\}$, $\mathcal{C}^*$ solves agent $i$'s problem,
\begin{align}
    &\max_{\mathcal{C}}~\bbE_{\mu(\mathcal{C^*})} \sum_{t=0}^{\infty}\beta^tu(s^i_{t}, c^i_{t}),\\
    &\text{s.t. }\eqref{eq:state_general_dynamics}\eqref{eq:general_borrowing_constraint}\eqref{eq:exo_general_dynamics} \text{ hold,  } c^i_{t}=\mathcal{C}(s^i_{t}, X_t, \bS_t), \\
    &{\text{given }} c^j_{t}=\mathcal{C}^*(s^j_{t}, X_t, \bS_t), j=1,\dots,N, j\neq i.
\end{align}

Note that in contrast to the perturbation method in \cite{reiter2009solving}, which only computes the solution around the stationary equilibrium in the
absence of aggregate shocks, a global solution method seeks to find the solution according to the stationary distribution $\mu(\mathcal{C^*})$ of the economy, which may significantly differ from the stationary equilibrium without aggregate shocks \citep{kekre2020monetary, bhandari2021inequality}.

We now make two remarks about the general setup we present above.
\begin{remark}[Discrete time setup] Throughout this paper, we formulate HA models in discrete time. Founded upon the methodological framework of \cite*{achdou2017income}, the study of HA models in continuous time has received a great deal of interest in recent years. Compared to discrete time, a continuous time setup admits explicit expectation integral formulas (with respect to specific forms of idiosyncratic shocks) and efficient numerical algorithms for solving the corresponding systems of partial differential equations (PDEs). However, when there are aggregate shocks, the stochastic PDE systems~\citep{carmona2018probabilistic2} must be derived and solved in order to obtain the global solution, which is much more challenging.
Here we use a discrete time setup so that the problem is easier to solve and more general forms of shocks can be included.
\end{remark}

\begin{remark}[Infinite horizon]
We restrict the setup to the infinite horizon for the sake of concision when introducing the algorithm. As seen in Section \ref{sec:general_algo}, our method can also handle finite horizon problems such as life cycle models with only minor modification.
\end{remark}

\subsection{Representation of the Agent Distribution and Generalized Moments}
The exposition of DeepHAM comprises two steps. The first is the introduction of generalized moments to replace the full agent distribution. This can be considered a model reduction step.
The second is the introduction of an algorithm to solve the reduced model.
We remark that the first step is of independent interest: the reduced model itself is a reliable and interpretable model that can be used as a general starting point for performing economic analysis. We discuss the idea of the first step in this subsection and present the detailed algorithm in the next subsection.

In HA models, a key question is what variables should be used to represent the whole economy. In algorithmic terms, these are what should be fed into the policy and value function approximators as input. 
Clearly, the individual state $s^i_t$ and the aggregate state variable $X_t$ should be taken into account.
The main question is therefore how to represent the empirical distribution $\bS_t=\{s^1_t,s^2_t,\dots,s^N_t\}$, which also affects the dynamics of $s^i_t$ and $X_t$. On this point, the existing literature generally uses one of the following two
approaches.

The first approach uses the vector $S_t$ \citep[see, e.g.,][]{maliar2021deep}, the 
full information of the distribution, to characterize the optimal policy and value functions.
However, there are two caveats. First, the dimension of $S_t$ is proportional to $N$. It is thus extremely expensive to deal with economies with a large number of agents. Second, the agent's optimal policy and value function should be invariant to the ordering of other agents' states. A function approximation form taking $S_t$ as the direct input cannot straightforwardly impose this restriction and must inefficiently process the irrelevant information of the agents' ordering.

The second approach uses finite moments, usually the first moment of $\bS_t$. This is the approach adopted by the KS method. It overcomes the two caveats discussed above of using the full vector $S_t$. However, the moments chosen may not carry the full information necessary for an agent to evaluate her current environment and act optimally. Under this simplification, the solution may deviate from the ground truth, especially in complex HA models.


To go beyond the above limitations, we introduce a class of generalized aggregate variables $Q_t\in\RR^{d_Q}$ into the state vector:
\begin{equation*}
    Q_t = \frac1N \sum_{i=1}^N \cQ(s^i_{t}).
\end{equation*}
Here the basis function $\cQ$ may take a pre-specified functional form, or it may be a general basis function with variational parameters. For pre-specified functional forms, for example, it could be the identity function, making $Q_t$ the first moment. $\cQ$ might also be an indicator function of whether an agent is at the borrowing constraint, so that $Q_t$ captures the share of hand-to-mouth agents \citep*{kaplan2014wealthy} in the economy. 
A $Q_t$ based on pre-specified $\cQ$ nests the moment representation we discuss above. A general basis function $\cQ$ can be parameterized with neural networks~\citep[see, e.g.,][]{han2019uniformly} and the optimal representation solved for in the algorithm. When $\cQ$ is a general basis function, we call the components of the resulting $Q_t$ \textit{``generalized moments''}.\footnote{A brief mathematical introduction to neural networks is presented in Appendix \ref{app:nn}.}

With $Q_t$ as the representation of the distribution, we use $(s^i_{t}, X_t, Q_t)$ as the state vector taken as input by
the policy and value function approximations. Conceptually, one can think of this in the following two ways.  %

First, instead of parameterizing the mapping $\left[(s^i_t, X_t, \bS_t)\mapsto \text{ output} \right]$ with neural network models,  we decompose it into $\left[(s^i_t, X_t, \bS_t)\mapsto (s^i_t, X_t, Q_t) \mapsto \text{ output} \right]$ and parameterize the two components by two neural networks, the first step is an encoding network and the second
step is a fitting network. In this sense, by specifying the dimension $d_Q$, we ensure that the complexity of the neural networks does not increase 
rapidly as $N$ increases. Furthermore, the final policy functions are permutation invariant by design. 
In this way, both shortcomings mentioned above are overcome.\footnote{Such a two-step decomposition through generalized moments is also closely linked to the recent literature \citep{zaheer2017deep} of approximating permutation invariant functions through deep neural networks.}

Second, $Q_t$ shares a similar representation to $O_t$, and can be interpreted as a vector of generalized moments. 
If $\cQ$ is parameterized by a set of specific functional forms informed by structural details, $Q_t$ are selected from a large set of interpretable aggregate moments. If $\cQ$ is directly parameterized by neural networks, optimizing $\cQ$ can guide the agent to finding the generalized moments $Q_t$ most relevant to their decision making. Compared to the KS method, generalized moments have the flexibility to more closely capture those features of the whole economy most relevant to the optimal decision rules, obtaining a more accurate macroeconomic model without sacrificing interpretability. In this regard, the generalized moments can be viewed as a set of ``numerically determined sufficient statistics'' of the model, which is the numerical counterpart of those analytical sufficient statistics we commonly see in structural models \citep{chetty2009sufficient, auclert2019monetary}.

In this paper, we use two separate sets of generalized moments, $Q^c_t$ and  $Q^V_t$, to extract the distribution information for the policy and value functions, respectively. 
This choice simplifies their updating rule.
Algorithmically, when we fed $Q^c_t/Q^V_t$ into the policy/value functions,
the variational parameters of the basis and policy/value parameters can be trained jointly end-to-end through the corresponding objective function. Using shared generalized moments for both the policy and value functions will be investigated in future work.

\subsection{Solution Method for Competitive Equilibrium}\label{sec:general_algo}
We first describe the algorithm for solving for the competitive equilibrium of economies of the form described in Section~\ref{sec:game_setup}. The algorithm for solving the constrained efficiency problem is quite similar and will be discussed in Section \ref{sec:general_constrained}. To solve the model, we rewrite agents' objective function based on the dynamic programming principle over $T$ periods. In the competitive equilibrium, for $\forall i \in \{1, \dots, N\}$, the policy function $\mathcal{C}^*$ solves agent $i$'s problem,
\begin{align}
\label{eq:unroll_bellman}
    \max_{\mathcal{C}}~&\bbE_{\mu(\mathcal{C}^*)} \left[\sum_{t=0}^{T} \beta^{t} u\left(s^i_{t}, c^i_{t}\right) + \beta^T V(s^i_{T}, X_T, \bS_T) \right],\\
    &\text{s.t. }\eqref{eq:state_general_dynamics}\eqref{eq:general_borrowing_constraint}\eqref{eq:exo_general_dynamics} \text{ hold,  } c^i_{t}=\mathcal{C}(s^i_{t}, X_t, \bS_t), \\
    &{\text{given }} c^j_{t}=\mathcal{C}^*(s^j_{t}, X_t, \bS_t), j=1,\dots,N, j\neq i,
\end{align}
where the value function $V$ is defined by
\begin{align}
\label{eq:value}
    V(s^i_{0}, X_0, \bS_0) =~&\bbE\left[\sum_{t=0}^{\infty} \beta^{t} u\left(s^i_{t}, c^i_{t}\right) \big| X_0, \bS_0 \right],\\
    &\text{s.t. }\eqref{eq:state_general_dynamics}\eqref{eq:general_borrowing_constraint}\eqref{eq:exo_general_dynamics} \text{ hold,  } c^i_{t}=\mathcal{C}^*(s^i_{t}, X_t, \bS_t), i=1,\dots,N.
\end{align}
The overall idea of DeepHAM is an iterative procedure starting from the initial guess of the policy $\mathcal{C}_0$, as presented in Algorithm~\ref{alg:game}. 
Each iteration includes three steps that we will discuss in detail: (a) prepare the stationary distribution $\mu(\mathcal{C})$; (b) update the value function according to \eqref{eq:value}; (c) optimize the policy function according to \eqref{eq:unroll_bellman}. We refer to one such high-level iteration step as a \textit{round} and repeat $N_k$ rounds until convergence. It is similar to the conventional value function iteration algorithm \cite[Chapter 3]{ljungqvist2018recursive}, except for the following two aspects. 

First, we parameterize both value and policy functions with neural networks, each of which nests two sub-networks with a feedforward architecture: one approximates the basis function $\cQ$, the other approximates the mapping from $(s^i_{t}, X_t, Q_t)$ to policy or value function values.\footnote{If we choose to use some pre-specified basis to define $Q$, such as the first moment, we will only have the second sub-network in both the policy and value functions.}
     
Second, we solve for optimal policy to maximize the total utility in \eqref{eq:unroll_bellman} over Monte Carlo simulations for $T$ periods, instead of one period, which is typically used in the conventional value function iteration algorithm. 
When the state vector is high dimensional, it is computationally expensive or even infeasible to update the policy with one period calculation in the whole state space. Instead, we update the parameters of the policy function neural network to maximize the expected total utility in \eqref{eq:unroll_bellman} over simulated paths. We choose $T > 1$ such that the error in the value function, which is discounted by $\beta^T$, will have little impact on the policy function optimization.
     

\begin{algorithm}
    \caption{DeepHAM for solving the competitive equilibrium}
  \begin{algorithmic}[1]
      \Require{Input:} the initial policy $\mathcal{C}_0$, the initial value and policy neural networks with parameters $\Theta^V$ and $\Theta^C$, respectively
     \For{$k=1,2,\dots,N_k$}
          \State prepare the stationary distribution $\mu(\mathcal{C}_{k-1})$ according to the policy $\mathcal{C}_{k-1}$
          \For{$m=1,2,\dots,N_{m_1}$}  \Comment{update the value function}
              \State sample $N_{b_1}$ samples of $(X_0, \bS_0)$ from $\mu(\cC_{k-1})$
              \State compute the realized total utility in \eqref{eq:realized_utility_game} through a single simulated path
              \State use the empirical version of \eqref{eq:value_objective_game} to compute the gradient $\nabla_{\Theta_V}$ \label{algline_gamevalue}
              \State update $\Theta^V$ with $\nabla_{\Theta_V}$
          \EndFor
          \For{$m=1,2,\dots,N_{m_2}$} \Comment{optimize the policy function}
              \State sample $N_{b_2}$ samples of $(X_0, \bS_0)$ from $\mu(\cC_{k-1})$
              \State use the empirical version of \eqref{eq:policy_objective_game} to compute the gradient $\nabla_{\Theta_C}$ \label{algline_gamepolicy}
              \State update $\Theta^C$ with $\nabla_{\Theta_C}$
          \EndFor
          \State define $\cC_k$ according to \eqref{eq:policy_form}
     \EndFor
  \end{algorithmic} 
  \label{alg:game}
\end{algorithm} 
Now we further explain the details of the three main steps of DeepHAM in the $k$-th round.

\paragraph{Preparing the stationary distribution} We simulate the economy~\eqref{eq:state_general_dynamics}\eqref{eq:exo_general_dynamics} forward for sufficiently many periods under the policy $\mathcal{C}_{k-1}$ to find the stationary distribution $\mu(\mathcal{C}_{k-1})$ of the economy. Then we store enough samples of $(X_0, \bS_0)$ according to $\mu(\mathcal{C}_{k-1})$, which will be used as the initial condition for later updating of the value and policy functions.

\paragraph{Updating the value function} Given the policy $\mathcal{C}_{k-1}$, updating the value function can be formulated as a supervised learning problem. Denote the parameters in the value function neural network by $\Theta^V = (\Theta^{VQ}, \Theta^{VO})$, where $\Theta^{VQ}$ are the parameters in the general basis function defining $Q^V_t$ and $\Theta^{VO}$ are the parameters in the function that maps $(s^i_{t}, X_t, Q^V_t)$ to value outcomes.
Our approximation to the value function can thus be written,
\begin{equation}
    V_{\text{NN}}(s^i_{t}, X_t, \bS_t; \Theta^V) \coloneqq \widetilde{V}_{\text{NN}}(s^i_{t}, X_t, Q^V_t; \Theta^{VO}) =
    \widetilde{V}_{\text{NN}}(s^i_{t}, X_t, \frac1N \sum_{i=1}^N \cQ_{\text{NN}}(s^i_{t}; \Theta^{VQ})); \Theta^{VO}).
\label{eq:vnn_game_formula}
\end{equation}
We want to use $V_{\text{NN}}$ to approximate the agents' expected lifetime utility under the policy $\mathcal{C}_{k-1}$, i.e.,
\begin{equation}
\label{eq:vnn_game}
 V_{\text{NN}}(s^i_{t}, X_t, \bS_t; \Theta^V) \approx \bbE \left[\sum_{\tau=0}^{\infty}\beta^{\tau}u(s^i_{t+\tau},c^i_{t+\tau}) \,\Big |\,s^i_{t}, X_t, \bS_t \right].
\end{equation}
However, evaluating the expectation in \eqref{eq:vnn_game} is still computationally expensive. To reduce the computational cost, given each sample $(s^i_{0}, X_0, \bS_0)$ from the stationary distribution $\mu(\cC_{k-1})$, we only simulate a single path for $T_\text{simul}$ periods (with $T_\text{simul}$ sufficiently large) under the policy $\cC_{k-1}$ to get the truncated realized total utility
\begin{equation}
\label{eq:realized_utility_game}
    \widehat{V}^i=\sum_{\tau=0}^{T_\text{simul}}\beta^{\tau}u(s^i_{\tau},c^i_{\tau}).
\end{equation}
Note that $\widehat{V}^i_t$ is a random variable influenced by the realization of the idiosyncratic and aggregate shocks. Still, we know that the true value function minimizes the expected difference with the realized total utility. 
Thus, we only need to solve the following regression problem to update the value function:
\begin{equation}
\label{eq:value_objective_game}
    \min_{\Theta^{V}} \bbE_{\mu(\cC_{k-1})} \left[V_{\text{NN}}(s^i_{0}, X_0, \bS_0; \Theta^V) - \widehat{V}^i\right]^2.
\end{equation}
We use the stochastic gradient descent algorithm to solve \eqref{eq:value_objective_game}. 
Specifically, in each update step, we sample $N_{b_1}$ samples of $(X_0, \bS_0)$ from $\mu(\cC_{k-1})$, use the empirical version of \eqref{eq:value_objective_game} to compute the gradient with respect to $\Theta^V$ by backpropagation,
and update $\Theta^V$ accordingly.
We repeat $N_{m_1}$ steps to achieve convergence.
As $\Theta^V = (\Theta^{VQ}, \Theta^{VO})$, we also obtain, at the end of the update, the updated basis function $\cQ_\text{NN}(\cdot; \Theta^{VQ})$ and the generalized moments $Q_t$ at the same time.\footnote{Backpropagation is an algorithm allowing an efficient computation of all partial derivatives of the neural network (composition of a series of functions) with respect to its parameters.}

\paragraph{Optimizing the policy function} In the competitive equilibrium, the policy function is updated iteratively following the spirit of fictitious play \citep{brown1951iterative}. A similar idea has been used in \cite{han2020deep} and \cite{hu2021deep} to solve stochastic differential games based on neural networks.
Similar to our approach to the value function, we will update the parameters associated with the policy function neural network through stochastic gradient descent. 
We call each update a ``play''.
In each ``play'', we fix everyone but agent $i = 1$'s policy as that from the last play, and consider agent $i = 1$'s utility maximization problem to update the neural network parameters, to get the new policy in this ``play''. All agents then adopt the new policy in this ``play''. We repeat the ``plays'' until convergence.\footnote{In a more general setup, we can also fix the other agents' policies for several ``plays" and then update from the agent $i=1$'s policy.}

\begin{figure}[h!]
\centering
  \includegraphics[width=0.95\textwidth]{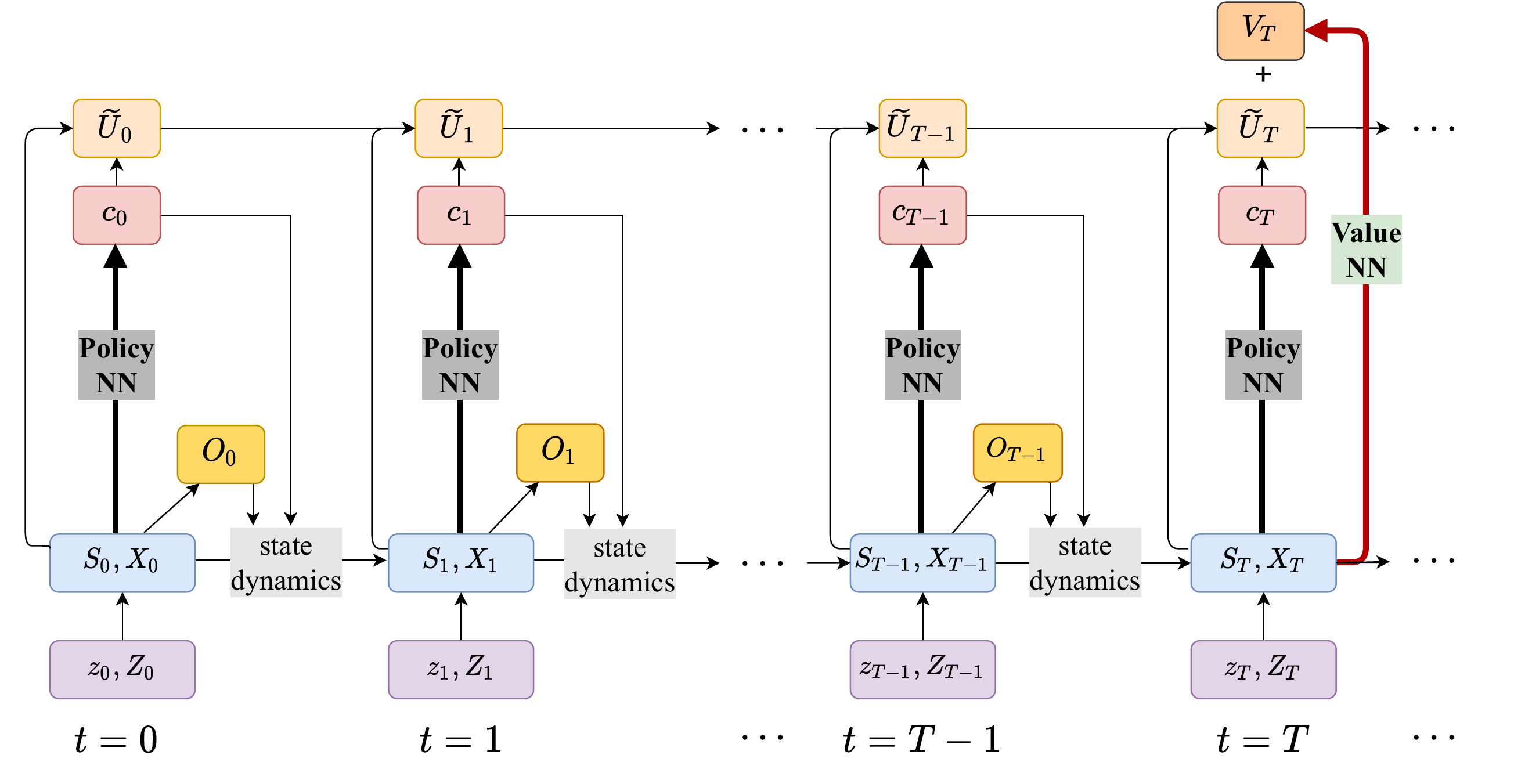}
  \caption{Computational graph to solve general HA models using DeepHAM. $S_t$, $z_t$, and $c_t$ denote the collection of all agents' states, idiosyncratic shocks, and decisions at time $t$, respectively. $Z_t$ denotes aggregate shocks at time $t$. $\widetilde{U}_t$ denotes the collection of all agents' cumulative utilities up to period $t$, i.e., $\widetilde{U}^i_{t} =  \sum_{\tau=0}^{t} \beta^{\tau} u\left(s^i_{\tau}, c^i_{\tau}\right).$}
  \label{fig:general_graph}
\end{figure}

For the utility maximization problem of agent $i=1$, the algorithm essentially builds upon the one proposed in \cite{han2016deep}: optimizing the parameters of the policy function neural network over simulated paths. Given the updated value function $V_{\text{NN}}(s^i_{t}, X_t, \bS_t; \Theta^V)$ in the same round and other agents' policy function from the last ``play'', agent $i = 1$ aims to solve
\begin{equation}
\label{eq:policy_objective_game}
    \max_{\Theta^C}\mathbb{E}_{\mu(\cC_{k-1})} \left[\sum_{t=0}^{T} \beta^{t} u(s^i_{t}, c^i_{t}) + \beta^T V_{\text{NN}}(s^i_{T}, X_T, \bS_T; \Theta^V)\right],
\end{equation}
with her policy parameterized by neural networks in the form of
\begin{align}
    c^i_{t} & = \mathcal{C}(s^i_{t}, X_t, \bS_t; \Theta^C) \notag \\
    & = \left(h_u(s^i_{t}, X_t, \bS_t) - h_l(s^i_{t}, X_t, \bS_t)\right) \odot c_{\text{NN}}\left(s^i_{t}, X_t, \bS_t; \Theta^C\right) + h_l(s^i_{t}, X_t, \bS_t). \label{eq:policy_form}
\end{align}
Here the outputs of $h_u(\cdot), h_l(\cdot)$, and $c_{\text{NN}}(\cdot)$ are $d^c$ dimensional, and $\odot$ denotes element-wise multiplication. We use the sigmoid function $\frac{1}{1+e^{-x}} \in [0,1]$ as the last composed function of $c_{\text{NN}}(\cdot)$, so that the inequality constraints~\eqref{eq:general_borrowing_constraint} are always satisfied. For a mathematical introduction to the composition structure of neural networks, see Appendix \ref{app:nn}.
Similar to the value function, the parameters in the policy function neural network $\Theta^C = (\Theta^{CQ}, \Theta^{CO})$, where $\Theta^{CQ}$ are parameters in the general basis function and $\Theta^{CO}$ are parameters in the function that maps $(s^i_{t}, X_t, Q_t)$ to policy outcomes.
So we have
\begin{equation}
    c_{\text{NN}}(s^i_{t}, X_t, \bS_t; \Theta^C) = \widetilde{c}_{\text{NN}}(s^i_{t}, X_t, Q^C_t; \Theta^{CO}) =
    \widetilde{c}_{\text{NN}}(s^i_{t}, X_t, \frac1N \sum_{i=1}^N \cQ_{\text{NN}}(s^i_{t}; \Theta^{CQ})); \Theta^{CO}).
\label{eq:cnn_formula}
\end{equation}
Once more, we use the stochastic gradient descent algorithm corresponding to \eqref{eq:policy_objective_game}.
We sample $N_{b_2}$ samples of $(X_0, \bS_0)$ from $\mu(\cC_{k-1})$ as the initial conditions, use the empirical version of \eqref{eq:value_objective_game} to compute the gradient with respect to $\Theta^C$, and update $\Theta^C$ accordingly.
The gradient with respect to $\Theta^C$ can be obtained by backpropagation as well, because all the component functions in \eqref{eq:policy_objective_game}\eqref{eq:policy_form}\eqref{eq:cnn_formula} are explicit and differentiable; see Figure~\ref{fig:general_graph} for the computational graph corresponding to \eqref{eq:policy_objective_game}.

From the above description, we can see one merit of DeepHAM: its ready ability to handle models with aggregate shocks. The algorithm remains almost the same 
when solving models with aggregate shocks or without. In contrast, the continuous time PDE approach \citep{achdou2017income} can solve models without aggregate shocks efficiently (in the low-dimensional case), but faces challenges in the presence of aggregate shocks.

\subsection{Extension to Constrained Efficiency Problem}\label{sec:general_constrained}
In the constrained efficiency problem, a benevolent social planner seeks to find a policy rule $\mathcal{C}$, determining each agent's decision variable $c^i_{t}$, in order to maximize the discounted sum of social welfare $\Omega(\barbS_t)$. The social welfare depends on the collection of all agents' state-control pairs:
\begin{align}
    &\max_{\mathcal{C}}~ \bbE_{\mu(\mathcal{C})} \sum_{t=0}^{\infty}\beta^t \Omega(\barbS_t),\\
    &\text{s.t. }\eqref{eq:state_general_dynamics}\eqref{eq:general_borrowing_constraint}\eqref{eq:exo_general_dynamics} \text{ hold,  } c^i_{t}=\mathcal{C}(s^i_{t}, X_t, \bS_t), i=1,\dots,N.
\end{align}
Here the social welfare function can take the utilitarian form $\Omega(\barbS_t) =  \frac1N\sum_{i=1}^N u(s^i_{t},c^i_{t})$, or $\Omega(\barbS_t) =  \sum_{i=1}^N \omega_i u(s^i_{t},c^i_{t})$ with Negishi weights $\omega_i = \frac{u_c(s^i_{t},c^i_{t})}{\sum_{i = 1}^N u_c(s^i_{t},c^i_{t})}$ \citep{bhandari2021inequality}, or other general forms.

The overall procedure for solving the constrained efficiency problem is the same as that for solving the competitive equilibrium, as presented in Algorithm~\ref{alg:sp}: each round consists of (a) preparing the stationary distribution, (b) updating the value function, and (c) optimizing the policy function. Below, we explain the three steps in the $k$-th round in detail and mainly highlight the differences between these and the procedure for solving for the value and policy objectives in competitive equilibrium.

\begin{algorithm}[!ht]
    \caption{DeepHAM for solving the constrained efficiency problem}
  \begin{algorithmic}[1]
      \Require{Input:} the initial policy $\mathcal{C}_0$, the initial value and policy neural networks with parameters $\Theta^V$ and $\Theta^C$, respectively
     \For{$k=1,2,\dots,N_k$}
          \State prepare the stationary distribution $\mu(\mathcal{C}_{k-1})$ according to the policy $\mathcal{C}_{k-1}$
          \For{$m=1,2,\dots,N_{m_1}$}  \Comment{update the value function}
              \State sample $N_{b_1}$ samples of $(X_0, \bS_0)$ from $\mu(\cC_{k-1})$
              \State compute the realized total social welfare in \eqref{eq:realized_utility_sp} through a single simulated path
              \State use the empirical version of \eqref{eq:value_objective_sp} to compute the gradient $\nabla_{\Theta_V}$
              \State update $\Theta^V$ with $\nabla_{\Theta_V}$
          \EndFor
          \For{$m=1,2,\dots,N_{m_2}$} \Comment{optimize the policy function}
              \State sample $N_{b_2}$ samples of $(X_0, \bS_0)$ from $\mu(\cC_{k-1})$
              \State use the empirical version of \eqref{eq:policy_objective_sp} to compute the gradient $\nabla_{\Theta_C}$
              \State update $\Theta^C$ with $\nabla_{\Theta_C}$
          \EndFor
          \State define $\cC_k$ according to \eqref{eq:policy_form}
     \EndFor
  \end{algorithmic} 
  \label{alg:sp}
\end{algorithm}

\paragraph{Preparing the stationary distribution} This is done exactly the same as for the competitive equilibrium problem. We simulate the economy~\eqref{eq:state_general_dynamics}\eqref{eq:exo_general_dynamics} forward for sufficiently many periods under the policy $\mathcal{C}_{k-1}$ to find the stationary distribution $\mu(\mathcal{C}_{k-1})$ of the economy. Then we store enough samples of $(X_0, \bS_0)$ according to $\mu(\mathcal{C}_{k-1})$, which will be used as the initial condition for later updating of the value and policy functions.

\paragraph{Updating the value function} Given the policy $\mathcal{C}_{k-1}$, we want to use $V_{\text{NN}}$ to approximate the expected total social welfare under the policy $\mathcal{C}_{k-1}$, i.e.,
\begin{equation}
\label{eq:vnn_sp}
 V_{\text{NN}}(X_t, \bS_t; \Theta^V) \approx \bbE \left[\sum_{\tau=0}^{\infty}\beta^{\tau}\Omega(\barbS_{t+\tau}) \,\Big |\ X_t, \bS_t \right],
\end{equation}
where 
\begin{equation}
    V_{\text{NN}}(X_t, \bS_t; \Theta^V) = V_{\text{NN}}(X_t, Q^V_t; \Theta^{VO}) =
    V_{\text{NN}}(X_t, \frac1N \sum_{i=1}^N \cQ_{\text{NN}}(s^i_{t}; \Theta^{VQ}); \Theta^{VO}).
\label{eq:vnn_sp_formula}
\end{equation}
Similarly, to avoid the computational cost of evaluating the expectation in \eqref{eq:vnn_sp} given each sample $(X_0, \bS_0)$ from the stationary distribution $\mu(\cC_{k-1})$, we only simulate a single path for $T_\text{simul}$ periods (with $T_\text{simul}$ sufficiently large) under the policy $\cC_{k-1}$ to obtain the truncated realized total social welfare
\begin{equation}
\label{eq:realized_utility_sp}
    \widehat{V}=\sum_{\tau=0}^{T_\text{simul}}\beta^{\tau}\Omega(\barbS_{\tau}).
\end{equation}
Then we only need to solve the following regression problem to update the value function:
\begin{equation}
\label{eq:value_objective_sp}
    \min_{\Theta^{V}} \bbE_{\mu(\cC_{k-1})} \left[V_{\text{NN}}(X_0, \bS_0; \Theta^V) - \widehat{V}\right]^2.
\end{equation}
This can be done using stochastic gradient descent in the same way as for \eqref{eq:value_objective_game}.

\paragraph{Optimizing the policy function} In order to find the constrained optimum, we need to update the policy function by solving
\begin{equation}
\label{eq:policy_objective_sp}
    \max_{\Theta^C}\mathbb{E}_{\mu(\cC_{k-1})} \left[\sum_{t=0}^{T} \beta^{t} \Omega(\barbS_{t}) + \beta^T V_{\text{NN}}(X_T, \bS_T; \Theta^V)\right].
\end{equation}
Compared to the policy function optimization in the competitive equilibrium problem, here we get rid of the fictitious play step and instead optimize all the agents' policies simultaneously to maximize the total social welfare. The optimization problem \eqref{eq:policy_objective_sp} can be solved in the same way as for the problem \eqref{eq:policy_objective_game} with the stochastic gradient descent algorithm.
The gradient with respect to $\Theta^C$ can be obtained by backpropagation in the same computational graph as in Figure~\ref{fig:general_graph}.\\

The above description demonstrates that the DeepHAM Algorithm \ref{alg:sp} for solving the constrained efficiency problem is identical to Algorithm \ref{alg:game} for the competitive equilibrium, except for the two differences in lines \ref{algline_gamevalue} and \ref{algline_gamepolicy}, in which we use different objectives for the value function and policy function corresponding to the different setting. Meanwhile, the pipeline of data sampling and optimization methods for these different objectives is exactly the same. In this sense, DeepHAM can solve the constrained efficiency problem as easily as the competitive equilibrium problem.

\section{DeepHAM for the Krusell-Smith Model}\label{sec:KS}
In this section, we illustrate DeepHAM on the classic Krusell-Smith model, and highlight the advantages of this method. 
\subsection{Model Setup}
The setup follows \cite{den2010comparison}. Household $i$'s state $s^i_{t}= (a^i_{t}, z^i_{t}) \in \RR^2$, with beginning-of-period wealth $a^i_{t}$, employment status $z^i_{t}\in \{0,1\}$. The consumption $c^i_{t}\in \RR$ is the control variable. Households have log utility over consumption. 
$Z_t\in \{Z^{h},Z^{l}\}$ denotes aggregate productivity.
The process $z^i_{t}, Z_t$ follows a first-order Markov process. 
The aggregate state variable $X_t = Z_t$, so its dynamics \eqref{eq:exo_general_dynamics} are trivial.
The state dynamics of the household comes from the household budget constraint,
\begin{align*}
    a^i_{t+1} &=(1+r_t-\delta)a^i_{t} + [(1-\tau_t)\bar{l}z^i_{t}+b(1-z^i_{t})]w_t- c^i_{t},\\
    a^i_{t+1} &\geq 0, \quad  c^i_{t} \geq 0,
\end{align*}
where the net rate of return of capital is $r_t-\delta$, with depreciation rate $\delta$.
The factor prices $r_t, w_t$ are determined by the first order condition (FOC) of the representative firm, which produces with a Cobb-Douglas technology $Y_t = Z_t K_t^\alpha L_t^{1-\alpha}$, in the competitive factor market,
\begin{align}
    w_t=Z_t(1-\alpha)(K_t/L_t)^\alpha, \quad r_t=Z_t\alpha(K_t/L_t)^{\alpha-1},
\end{align}
with aggregate capital $K_t =\frac1N\sum_{i=1}^N a^i_{t}$ and labor supply $L_t = \bar{l}(L^h\ONE_{Z_t=Z^h} + L^l\ONE_{Z_t=Z^l})$, in which $\bar{l}$ is the time endowment of each agent. Unemployed agents $(z^i_{t}=0)$ receive unemployment benefits $b w_t$ where $b$ is the unemployment benefit rate.
Employed agents $(z^i_{t}=1)$ earn after-tax labor income $(1-\tau_t)\bar{l}w_t$, with tax rate $\tau_t=b(1-L_t)/\bar{l}L_t$, such that government budget constraint always holds (total tax income equals unemployment benefits). 
This completes the specification of \eqref{eq:state_general_dynamics}. The borrowing constraint and non-negative consumption constraint specifies \eqref{eq:general_borrowing_constraint}. The calibration of the model follows \cite{den2010comparison} and is presented in Appendix \ref{app:KS}.

\subsection{Results}
\begin{figure}[h!]
\centering
  \includegraphics[width=0.95\textwidth]{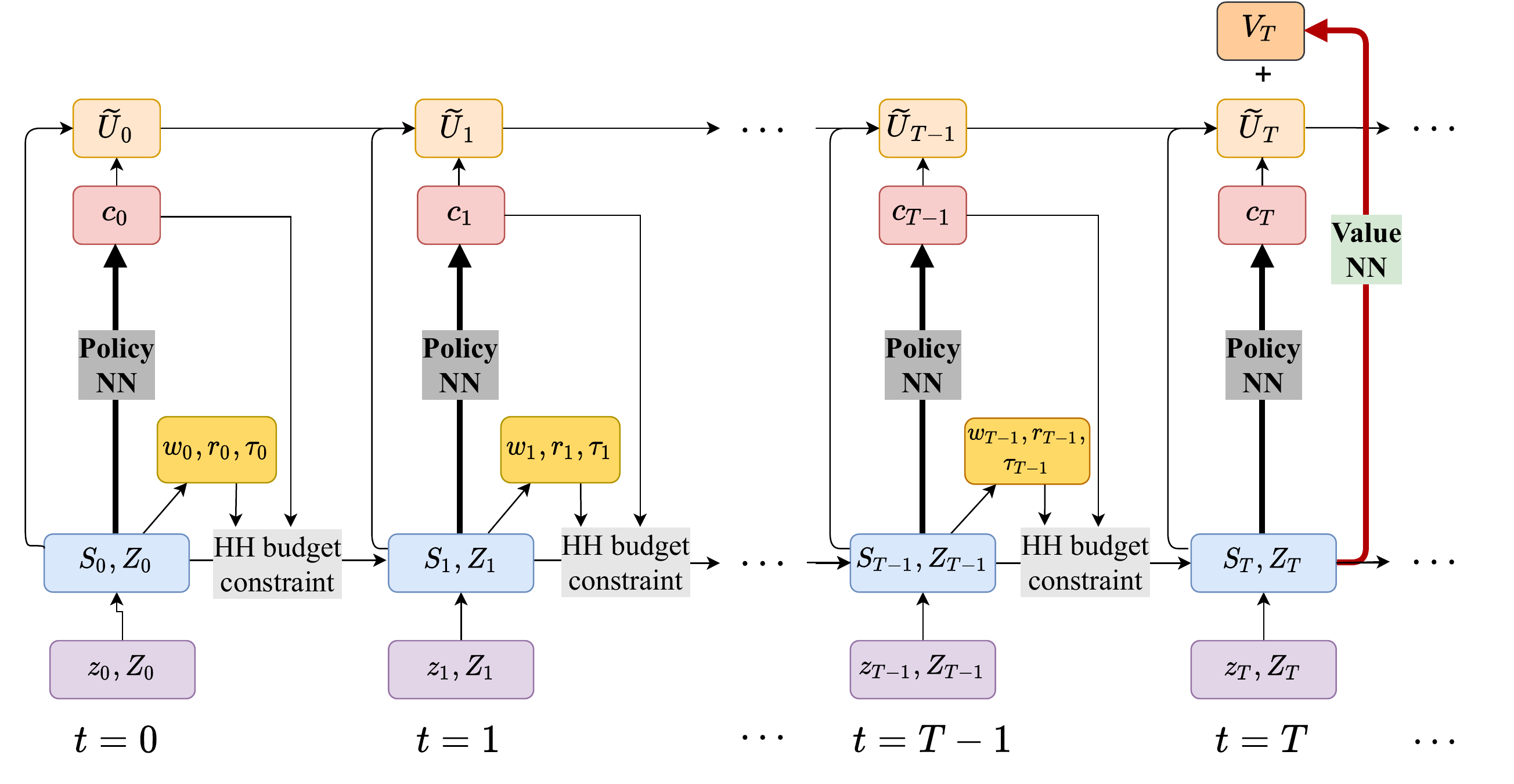}
  \caption{Computational graph to solve \cite{krusell1998income} using DeepHAM. $S_t$, $z_t$, $c_t$, and $\widetilde{U}_t$ denote the collection of all agents' states, idiosyncratic shocks, decisions, and cumulative utilities at time $t$, respectively. $Z_t$ denotes aggregate shocks at time $t$. Aggregate prices $w_t, r_t$ are determined by FOCs of the representative firm in the competitive factor market. Income tax rate $\tau_t$ depends on the aggregate shock $Z_t$ and is pinned down in the government budget constraint.}
  \label{fig:ks_graph}
\end{figure}

We solve the Krusell-Smith model described above in the case of $N=50$ using DeepHAM. We have tried other choices of $N = 100, 200, ...$, and we find $N = 50$ is large enough to approximate the solution to the Krusell-Smith model. The computational graph for this problem is shown in Figure \ref{fig:ks_graph}.

\subsubsection{Solution Accuracy}\label{sec:KS_acc}
In Table \ref{table:KS}, we compare the Bellman equation errors (defined in Appendix \ref{app:bellman}) of DeepHAM to the same error for the KS method implemented in \cite{maliar2010solving}. For DeepHAM, we present accuracy measures for the cases where we include (1) the first moment of the household wealth distribution and (2) one generalized moment of the household wealth distribution in the state variable. We present the standard deviation of the Bellman errors from multiple runs of the numerical algorithm in the last column of Table \ref{table:KS}.
We see that all results are statistically significant.\footnote{Following the moment construction in \cite{krusell1998income}, here the generalized moments are constructed on the wealth distribution, rather than the joint distribution of wealth and employment status.}
\begin{table}[!htb]  
\centering
\begin{tabular}{lcc}
\hline\hline
\multicolumn{1}{c}{Method and Moment Choice} & Bellman error  & Std of error \\ \hline
KS Method \citep{maliar2010solving}    & {0.0253}  & 0.0002\\ 
DeepHAM with 1st moment                    & {0.0184}  &  0.0023 \\ 
DeepHAM with 1 generalized moment        & 0.0151  & 0.0015 \\ 
\hline\hline
\end{tabular}
\caption{Comparison of solution accuracy for Krusell-Smith problem}
\label{table:KS}
\end{table}

As can be seen in Table \ref{table:KS}, the solutions obtained using DeepHAM are highly accurate. Compared to the KS method, DeepHAM with the first moment in the state vector reduces the Bellman equation error by 27.2\%. DeepHAM with one generalized moment reduces the error by 40.3\%.
Generalized moments play an important role in improving solution accuracy because they provide a more concise representation of the household distribution and extract more relevant information than the first moment. We discuss and interpret the generalized moment we obtain in the Krusell-Smith problem in the next subsection.\footnote{The KS method with the first moment is known to solve the Krusell-Smith model reasonably well. This is confirmed by the small Bellman error for the KS method in Table \ref{table:KS}. Though DeepHAM can further improve the solution accuracy, the simulated economy based on the DeepHAM solution is highly consistent with the simulation based on the KS method. We present this comparison in Appendix \ref{app:KS_compare}. This further confirms the accuracy of the DeepHAM solution for the Krusell-Smith model.}

\subsubsection{Generalized moments and redistributional effect}\label{sec:KS_gm}
\begin{figure}[!htb]
     \centering
     \begin{subfigure}[b]{0.45\textwidth}
         \centering
         \includegraphics[width=\textwidth]{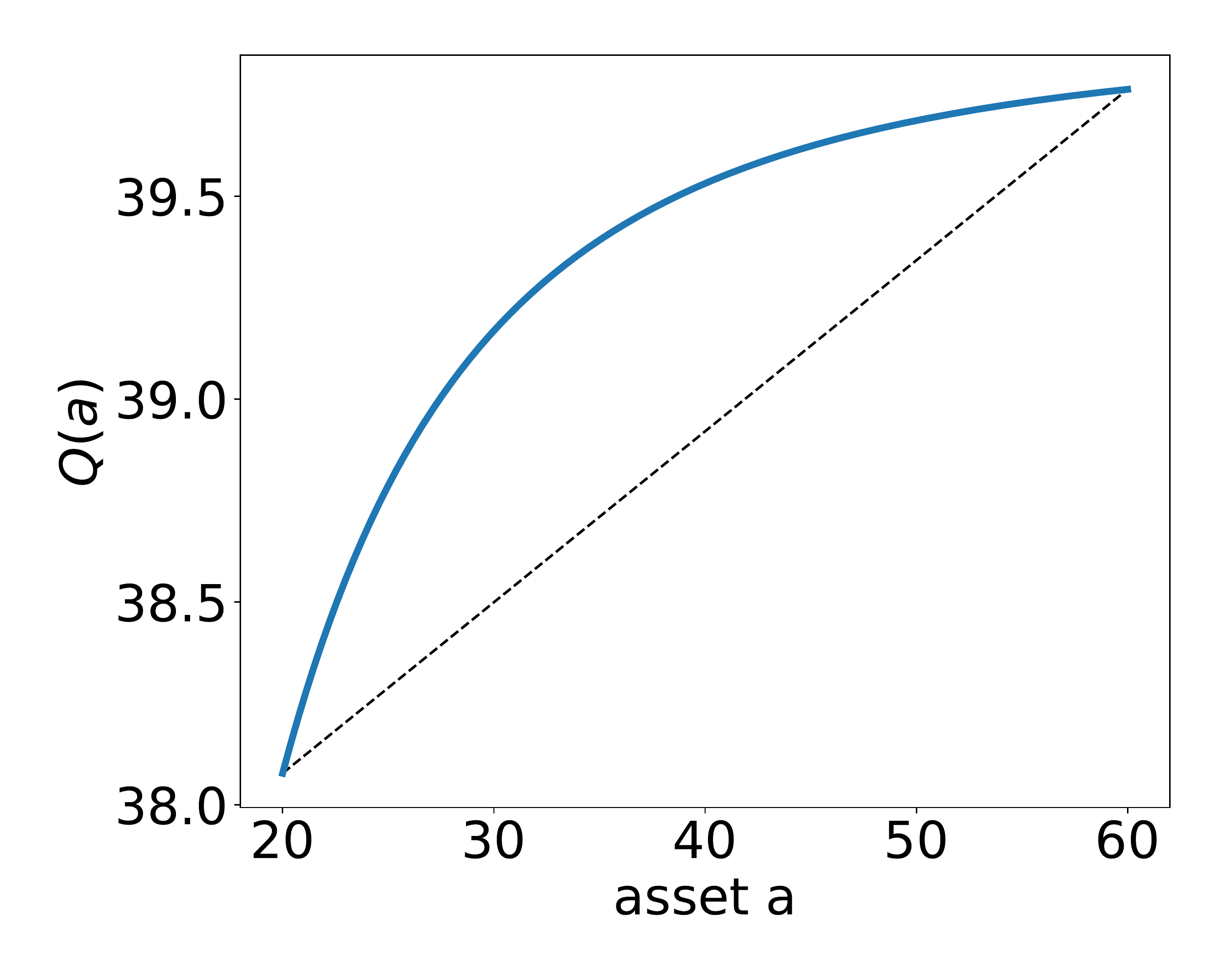}
         \caption{Plot of $\cQ_1(a)$}
         \label{fig:m1}
     \end{subfigure}
     \hfill
     \begin{subfigure}[b]{0.5\textwidth}
         \centering
         \includegraphics[width=\textwidth]{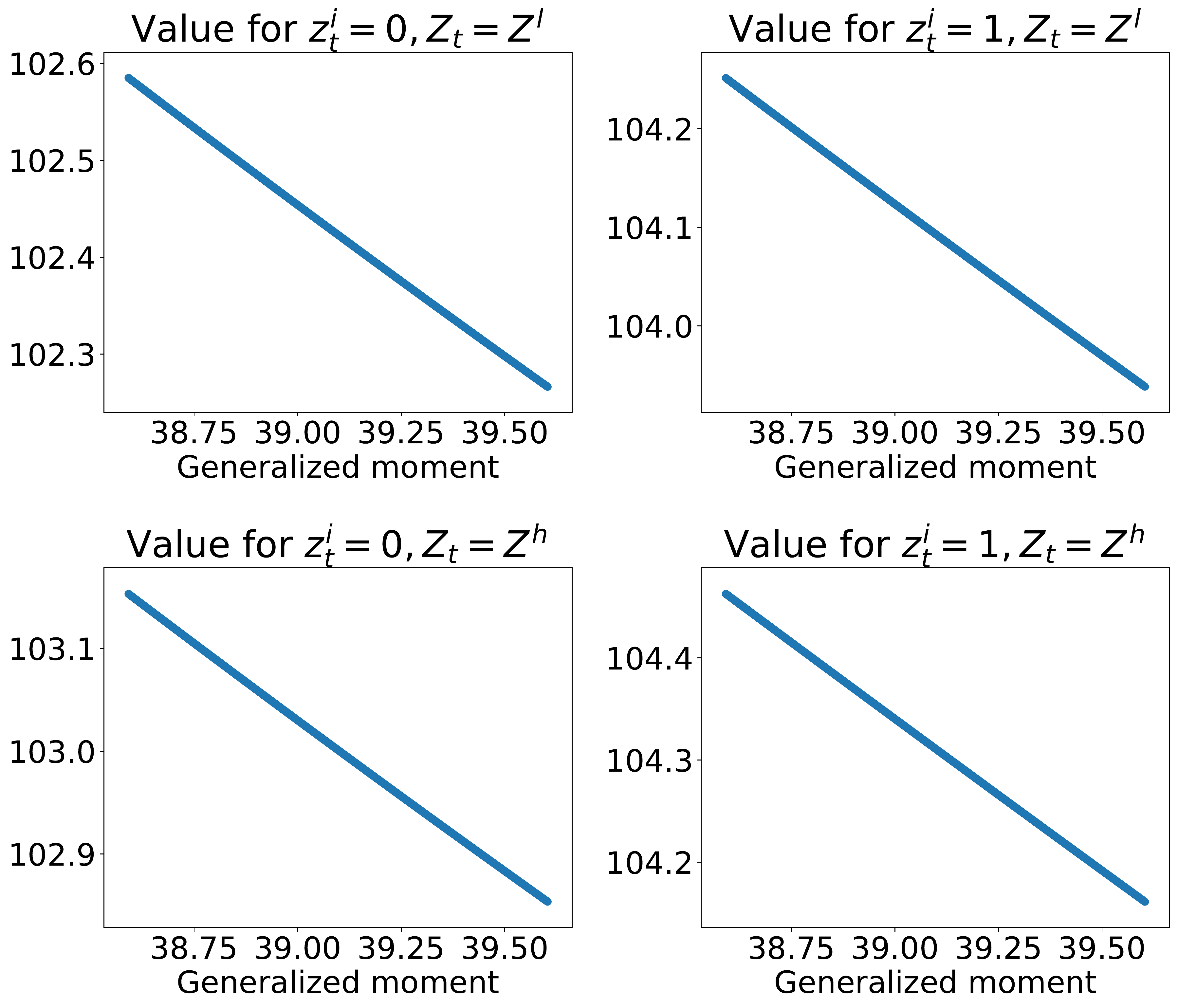}
         \caption{Map $\frac{1}{N}\sum_i \cQ_1(a_i)$ to the value function}
         \label{fig:m1v}
     \end{subfigure}
    \caption{Generalized moments\ for the Krusell-Smith problem. Left panel (solid blue line): a concave mapping from the individual asset to the basis function of the generalized moment. Right panel: mapping from the generalized moment to the value function, assuming households' individual assets fixed at the average level in the stationary equilibrium. Each figure in the right panel corresponds to one realization of idiosyncratic and aggregate shocks.}
        \label{fig:gm_KS}
\end{figure}

Among the three results in Table \ref{table:KS}, DeepHAM with one generalized moment yields the most accurate solution. To better understand the improvement, we visualize the mapping from individual asset holdings $a^i_{t}$ to the basis function $\cQ(a^i_{t})$, and the mapping from the generalized moment $\frac{1}{N}\sum_i \cQ(a^i_{t})$ to the value function in Figure \ref{fig:gm_KS}.

We find that the basis function is concave in the individual asset, while the value function is linear with regard to the generalized moment. That is, households with different levels of wealth will have heterogeneous contributions to the generalized moment: giving an additional unit of assets to poor households increases the generalized moment more than giving the same assets to rich households. This phenomenon means that a purely redistributional policy would affect aggregate welfare and dynamics even in the simple setup of \cite{krusell1998income}. Consider an unanticipated one-time policy shock (MIT shock): if one unit of the asset is redistributed from the richest households to the poorest households, the welfare of ``middle'' households who are not in the redistribution program would decrease on impact since the generalized moment increases. Such an unanticipated policy shock will lead to higher aggregate savings in the future, since there are fewer people on the borrowing constraint. The increase in savings would lead to a higher future wage and a lower future asset return. Under the calibration of this model, the ``middle'' households who are not in the redistribution program receive more capital income than labor income, so the unanticipated policy shock would make them worse off. This sensible logic differs from the implications of the solution of the KS method. According to the KS method, households' welfare only depends on the first moment and individual states. So the redistributional policy shock would have no instantaneous welfare impact on those ``middle'' households who are not in the redistribution program, since the first moment of individual wealth distribution would not change.

\section{DeepHAM for More Complex HA Models}\label{sec:jfv}
In this section, we use DeepHAM to solve a HA model with a financial sector and aggregate shocks, as proposed in \cite*{fernandez2019financial}. Compared to the Krusell-Smith model with the two-state aggregate shocks in Section \ref{sec:KS}, here the aggregate shocks take values in a continuous range, which makes the problem more costly to solve.

\subsection{\cite{fernandez2019financial}: Model Setup}
The setup is the discrete time version of \cite{fernandez2019financial}. We use the subscript $t$ and $t+\Delta t$ to highlight that the model comes from the discretization of a continuous time model, but should be interpreted as representing the dynamics between $t$ and $t+1$ in the general setup in Section \ref{sec:general_setup}. In this economy, there are $N$ households who save in risk-free bonds and consume. Their labor supply is exogenous and exposed to idiosyncratic shocks. There is a representative financial expert who issues risk-free bonds to households and invests in productive capital. A representative firm produces with capital from the financial expert and with labor supplied by the households. The growth rate of productive capital is exposed to aggregate shocks. 

\paragraph{Household's problem} For household $i$, her state is $s^i_t= (a^i_t, z^i_t)\in \RR^2$, with beginning-of-period risk-free asset $a^i_t$, and the idiosyncratic shocks on labor supply $z^i_t\in \{z_1,z_2\}$ with $0<z_1<z_2$. The process $z^i_t$ follows a first-order Markov process with ergodic mean 1 such that the aggregate labor supply $L_t = 1$. 
Household $i$ has constant relative risk aversion (CRRA) utility from consumption $c_t^i$ with parameter $\gamma >0$ and discount factor $e^{-\rho \Delta t}$.

The household budget and borrowing constraints determines the state dynamics \eqref{eq:state_general_dynamics} of household $i$: 
\begin{align}\label{eq:jfv_hhbc}
    a^i_{t+\Delta t} & = a^i_{t} + (w_{t} z^i_{t}+r_{t} a^i_{t}-c^i_{t})\Delta t,\\
    a^i_{t+\Delta t} & \geq 0, \quad c^i_{t} \geq 0,
\end{align}
where the aggregate prices are characterized below.
Aggregate risk-free asset demand $B_t =\frac1N\sum_{i=1}^N a^i_t.$ 

\paragraph{Representative firm's problem} 
The firm produces with Cobb-Douglas technology $Y_t = K_t^\alpha L_t^{1-\alpha}$. It hires labor $L_t$ from households at wage $w_t$, and rents capital $K_t$ from the financial expert at rental rate $rc_t$, both in the competitive factor market:
\begin{equation}\label{eq:jfv_wage}
    w_t = (1-\alpha){(K_t/L_t)}^{\alpha}, \quad rc_t = \alpha {(K_t/L_t)}^{\alpha-1}.
\end{equation}

\paragraph{Financial expert's problem} The representative financial expert issues a risk-free bond $B_t$ at rate $r_t$ to households, and rents capital $K_t$ at rate $rc_t$ to the representative firm. Her net worth $W_t = K_t - B_t$. For the financial expert, the instantaneous return rate on capital is exposed to aggregate shocks $Z_t$: 
$$\frac{K_{t+\Delta t} - K_t}{K_t} = (rc_t - \delta)\Delta t + \sigma Z_t\sqrt{\Delta t},$$
where $\delta$ is the depreciation rate of capital, $\sigma$ is the volatility of aggregate shocks, and $Z_t$ follows an i.i.d. standard normal distribution.\footnote{In the continuous time model, the aggregate shock is a white noise process with volatility $\sigma$. There would be a slight numerical difference to the discretized model, but the difference is tiny as we choose a tiny $\Delta t$.}

The financial expert has log utility with discount rate $\widehat{\rho} < \rho$ over consumption $\widehat{C}_t$, so she consumes a constant share of her net worth: $\widehat{C}_t = \widehat{\rho}W_t$, and chooses a leverage ratio proportional to excess return of risky capital $\frac{K_t}{W_t} = \frac{1}{\sigma^2}(rc_t - \delta - r_t)$. So the risk-free return is 
\begin{equation}\label{eq:jfv_r}
    r_t = \alpha {(K_t/L_t)}^{\alpha-1} - \delta - \sigma^2\frac{K_t}{W_t}.
\end{equation}

The budget constraint of the financial expert $W_{t + \Delta t}  = W_{t} + (rc_t - \delta) K_t\Delta t +\sigma K_{t} Z_{t} \sqrt{\Delta t} - B_t r_t \Delta t - \widehat{C}_t \Delta t$ implies the following dynamics of net worth $W_t$:
\begin{align}
\label{eq:JFV_exo_dynamics}
W_{t + \Delta t}  = W_{t} + \left(\alpha K_{t}^{\alpha-1}-\delta-\widehat{\rho}-\sigma^{2}\left(1-\frac{K_{t}}{W_{t}}\right) \frac{K_{t}}{W_{t}}\right) W_{t} \Delta t+\sigma K_{t} Z_{t} \sqrt{\Delta t}.
\end{align}
Using the general descriptive variables in Section \ref{sec:general_setup}, the aggregate state $X_t=W_t$. Since $K_t = B_t + W_t$, the evolution of $W_t$ only depends on $W_t, B_t =\frac1N\sum_{i=1}^N a^i_t$, and $Z_t$. The aggregate state dynamics \eqref{eq:exo_general_dynamics} are specified as \eqref{eq:JFV_exo_dynamics}. 
Equations \eqref{eq:jfv_hhbc}\eqref{eq:jfv_wage}\eqref{eq:jfv_r}, together with the stochastic process of $z^i_{t}$, complete the specification of \eqref{eq:state_general_dynamics} and \eqref{eq:general_borrowing_constraint}. The calibration of the model follows \cite{fernandez2019financial} and is presented in Appendix \ref{app:JFV}.

\subsection{Solution Accuracy and Efficiency}
We use DeepHAM to obtain the global solution to the problem described above with $N=50$. We compare the Bellman equation errors (see the definition in Appendix \ref{app:bellman}) of DeepHAM to the generalized KS method with the nonlinear perceived law of motion implemented in \cite{fernandez2019financial} in Table \ref{table:JFV_SSS}. For DeepHAM, we present accuracy measures for the cases where we include in the state variable (1) only the first moment or (2) one generalized moment of household asset distribution.

\begin{table}[!htb]  
\centering
\begin{tabular}{lcc}
\hline\hline
\multicolumn{1}{c}{Method and Moment Choice}  & Bellman error & Std of error \\ \hline
KS Method \citep{fernandez2019financial}                         & 0.00417       & 0.00011      \\
DeepHAM with 1st moment              & 0.00405       & 0.00059      \\ 
DeepHAM with 1 generalized   moment & 0.00422       &  0.00086         \\
\hline\hline
\end{tabular}
\caption{Comparison of solution accuracy on a HA model with a financial sector and aggregate shocks. The KS method refers to the solution method implemented by \citep{fernandez2019financial}.}
\label{table:JFV_SSS}
\end{table}
As we see in Table \ref{table:JFV_SSS}, the solutions obtained using DeepHAM are highly accurate. 
Compared to the generalized KS method with the nonlinear law of motion implemented by \cite{fernandez2019financial}, DeepHAM with either the first moment or a generalized moment can obtain global solutions with the same level of accuracy.
We present the standard deviation of the Bellman errors from multiple runs of the numerical algorithm in the last column of Table \ref{table:JFV_SSS}.\footnote{Note that \cite{fernandez2019financial} builds and solves the model in continuous time. To compare it with DeepHAM, which solves the discrete time version of the model, we evaluate their solution also on the discrete time Bellman equation error defined in Appendix \ref{app:bellman}. Since we chose a small $\Delta t$, the continuous time solution should give a meaningful approximate Bellman error in discrete time. We find that DeepHAM obtains an accurate solution which is comparable to \cite{fernandez2019financial}.}

Solving this HA model, with a financial sector and aggregate shocks which take values in a continuous range, takes DeepHAM 12\% longer than solving the simple Krusell-Smith model in Section \ref{sec:KS}. This result demonstrates the efficiency of DeepHAM in studying complex HA models with aggregate shocks: unlike the grid-based method, the computational cost of DeepHAM does not increase quickly when the number of state variables or grid points increases.

\section{DeepHAM for Constrained Efficiency Problem in HA Models} \label{sec:planner}
In this section, we solve the constrained efficiency problem in HA models using DeepHAM. In contrast to the competitive equilibrium, the constrained optimum of an HA model, defined as the allocation decided by a benevolent social planner who maximizes social welfare, is much harder to solve.
Existing literature only handles constrained optima of HA models without aggregate shocks \citep*{davila2012constrained, nuno2018social}, and at a much higher computational cost than for the competitive equilibrium of the same model.

In contrast, as presented in Section \ref{sec:general_constrained}, DeepHAM can solve the constrained efficiency problem as easily as it can solve the competitive equilibrium. 
We illustrate this advantage by using DeepHAM to solve the constrained efficiency problem in an Aiyagari model as in \cite{davila2012constrained}, and in a HA model with aggregate shocks.

\subsection{Model Setup}
\paragraph{Baseline setup without aggregate shocks} The baseline setup is an Aiyagari model that follows the ``high wealth dispersion'' calibration in \cite{davila2012constrained} to match empirical US wealth inequality. There are $N$ \textit{ex ante} homogeneous households in this economy. Household $i$'s labor supply is subject to idiosyncratic shocks $z^i_{t} \in \{e_0,  e_1, e_2\}$, which are i.i.d. across agents and follow a Markov process. Household $i$ accumulates asset $a^i_{t}$ in the form of real capital. 
Household $i$'s state $s^i_{t} = (a^i_{t}, z^i_{t})$ follows
\begin{align}\label{eq:davila_hhbc}
    a^i_{t+1} & =(1+r_t-\delta)a^i_{t} + w_t z^i_{t} - c^i_{t},\\
    a^i_{t+1} & \geq 0, \quad c^i_{t} \geq 0,
\end{align}
where consumption $c^i_{t}$ is the only control variable. The representative firm produces with a Cobb-Douglas technology $Y_t = K_t^\alpha L_t^{1-\alpha}$ and rents capital and hires labor in a competitive factor market. So the wage $w_t$ and capital rental rate $r_t$ are: 
\begin{align}\label{eq:davila_price}
    w_t=(1-\alpha)(K_t/L_t)^\alpha,\quad r_t=\alpha(K_t/L_t)^{\alpha-1},
\end{align}
where aggregate saving $K_t =\frac1N\sum_{i=1}^N a^i_{t}$ and labor supply $L_t= \bar{L}$ is constant.
This completes the specification of \eqref{eq:state_general_dynamics} and \eqref{eq:general_borrowing_constraint}. Since there is no aggregate shock in the baseline setup, there is no aggregate state variable $X_t$ nor are there dynamics as in \eqref{eq:exo_general_dynamics}.

The benevolent social planner seeks to find a policy rule $\mathcal{C}$ determining $c^i_{t}$ for all the households $i = 1, ..., N$ and $t = 0, 1, ..., \infty$ to maximize the utilitarian objective,
\begin{align}
    \max_{\mathcal{C}}~\frac1N\bbE_{\mu(\mathcal{C})} \sum_{i=1}^N \sum_{t=0}^{\infty}\beta^t u(c^i_{t}),
\end{align}
subject to the constraints in \eqref{eq:davila_hhbc}.\footnote{Given the form of the utilitarian objective, we have an alternative approach to approximate the expected social welfare besides the one presented in Section~\ref{sec:general_constrained}. We can learn the individual value function defined in \eqref{eq:vnn_game} with the approximation form \eqref{eq:vnn_game_formula} and use that to approximate the expected total social welfare by taking the average of the individual value function. We have used the latter approach in this work.}

\paragraph{Setup with aggregate shocks} We also solve the constrained efficiency problem of a HA model with aggregate shocks. On top of the baseline model above, we introduce aggregate productivity shocks $Z_t \in \{Z^l, Z^h\}$, that follow a Markov process, on the production technology of the representative firm $Y_t = Z_t K_t^\alpha L_t^{1-\alpha}$, such that the factor prices are,
\begin{align}\label{eq:davila_price2}
    w_t=Z_t(1-\alpha)(K_t/L_t)^\alpha,\quad r_t=Z_t\alpha(K_t/L_t)^{\alpha-1},
\end{align}
where aggregate saving $K_t =\frac1N\sum_{i=1}^N a^i_{t}$ and labor supply $L_t = (L^h\ONE_{Z_t=Z^h} + L^l\ONE_{Z_t=Z^l})$.  Using the general descriptive variables in Section \ref{sec:general_setup}, the aggregate state variable $X_t = Z_t$.
We also introduce countercyclical idiosyncratic risk to the model, so that the probability that households enter the low income state $z^i_{t} = e_0$ becomes larger in the bad aggregate state, and smaller in the good aggregate state. Our setup follows the ``integration principle'' proposed by \cite{krusell2009revisiting}, so that when aggregate shocks are eliminated, the model will exactly reduce to the baseline setup. Equations \eqref{eq:davila_hhbc} and \eqref{eq:davila_price} (or \eqref{eq:davila_price2}), together with the stochastic process of $z^i_{t}$, complete the specifications of \eqref{eq:state_general_dynamics} and \eqref{eq:general_borrowing_constraint}.  The calibration of both models are presented in Appendix \ref{app:Davila}.

\subsection{Results}
We solve the constrained planner's problems with $N=50$ in both the baseline model without aggregate shocks, and in the model with aggregate shocks and countercyclical idiosyncratic shocks. The equilibrium statistics of these problems are presented in Table \ref{table_Davila} and \ref{table_Davila2}. In comparison, we also present equilibrium statistics under the competitive equilibrium of the same models.\footnote{When solving the constrained efficiency problem, we usually find two local maxima of the problem. Since we are solving the constrained planner's problem, we only take the local optimum with higher expected total welfare. We leave the study of the ``second best'' constrained optimum for future research.}

\begin{table}[!htb]
\centering
\begin{tabular}{c|cc|cc}
\hline\hline& \multicolumn{2}{c|}{No aggregate shock} & \multicolumn{2}{c}{Aggregate shock}    \\ 
 &
  \multicolumn{1}{c}{Market} &
  Constrained Opt. &
  \multicolumn{1}{c}{Market} &
  Constrained Opt. \\ \hline
Average assets       & \multicolumn{1}{c}{30.635}  & 119.741  & \multicolumn{1}{c}{34.296}  & 95.811   \\ 
Output               & \multicolumn{1}{c}{10.294}  & 16.816   & \multicolumn{1}{c}{12.159}  & 17.592   \\ 
Capital-output ratio & \multicolumn{1}{c}{2.976}   & 7.120    & \multicolumn{1}{c}{2.821}   & 5.446    \\ 
Interest rate        & \multicolumn{1}{c}{4.097\%} & -2.944\% & \multicolumn{1}{c}{4.678\%} & -1.433\% \\ 
\begin{tabular}[c]{@{}c@{}}Coefficient of variation\\ of wealth\end{tabular} &
  \multicolumn{1}{c}{2.621} &
  2.483 &
  \multicolumn{1}{c}{2.574} &
  2.924 \\ 
Wealth Gini          & \multicolumn{1}{c}{0.864}   & 0.862    & \multicolumn{1}{c}{0.812}   & 0.878    \\ 
\begin{tabular}[c]{@{}c@{}}Coefficient of variation\\ of consumption\end{tabular} &
  \multicolumn{1}{c}{1.548} &
  0.710 &
  \multicolumn{1}{c}{1.699} &
  0.736 \\ 
Consumption Gini     & \multicolumn{1}{c}{0.615}   & 0.386    & \multicolumn{1}{c}{0.578}   & 0.388    \\\hline\hline
\end{tabular}
\caption{Equilibrium statistics in the market outcome (competitive equilibrium) and constrained optimum for models without or with aggregate shocks. }
\label{table_Davila}
\end{table}

\begin{table}[!htb]
\centering
\begin{tabular}{c|cc|cc}
\hline\hline & \multicolumn{2}{c|}{Positive aggregate   shock} & \multicolumn{2}{c}{Negative aggregate   shock} \\ 
 &
  \multicolumn{1}{c}{Market} &
  Constrained Opt. &
  \multicolumn{1}{c}{Market} &
  Constrained Opt. \\ \hline
Average assets       & \multicolumn{1}{c}{36.316}      & 99.793       & \multicolumn{1}{c}{32.260}      & 91.826       \\ 
Output               & \multicolumn{1}{c}{13.925}      & 20.038       & \multicolumn{1}{c}{10.393}      & 15.146       \\ 
Capital-output ratio & \multicolumn{1}{c}{2.608}       & 4.980        & \multicolumn{1}{c}{3.104}       & 6.063        \\ 
Interest rate        & \multicolumn{1}{c}{5.147\%}     & -1.116\%     & \multicolumn{1}{c}{4.208\%}     & -1.750\%     \\ 
\begin{tabular}[c]{@{}c@{}}Coefficient of variation\\ of wealth\end{tabular} &
  \multicolumn{1}{c}{2.533} &
  2.894 &
  \multicolumn{1}{c}{2.614} &
  2.953 \\ 
Wealth Gini          & \multicolumn{1}{c}{0.815}       & 0.877        & \multicolumn{1}{c}{0.805}       & 0.877        \\ 
\begin{tabular}[c]{@{}c@{}}Coefficient of variation\\ of consumption\end{tabular} &
  \multicolumn{1}{c}{1.693} &
  0.756 &
  \multicolumn{1}{c}{1.697} &
  0.713 \\ 
Consumption Gini     & \multicolumn{1}{c}{0.599}       & 0.407        & \multicolumn{1}{c}{0.542}       & 0.345         \\ \hline\hline
\end{tabular}
\caption{Equilibrium statistics in the market outcome (competitive equilibrium) and constrained optimum for the HA model with aggregate shocks, conditional on the realization of aggregate shock.}
\label{table_Davila2}
\end{table}
The main findings are as follows. First, in both models (with or without aggregate shocks), the constrained optimum requires a much higher level of capital than the competitive equilibrium. In the absence of aggregate shocks, the planner chooses a capital level 3.90 times that of the laissez-faire equilibrium, which is consistent with the finding of \cite{davila2012constrained}. This is because the planner with the utilitarian objective aims to redistribute from rich households to poor households in order to improve social welfare. Since poor households have a higher labor income share, the planner would raise the aggregate capital level, so that the wage rate increases and capital return decreases according to equations \eqref{eq:davila_price} and \eqref{eq:davila_price2}. By raising the aggregate capital level, such a redistribution leaves poor households better off. Meanwhile, in both models, the constrained efficiency problem features a similar level of wealth inequality and a lower level of consumption inequality relative to the market outcome.

Second, compared to the constrained optimum in the absence of aggregate shocks, the model with aggregate shocks features a lower level of aggregate capital stock. With aggregate shocks and countercyclical unemployment risks, the planner chooses a capital level 2.79 times that of the laissez-faire equilibrium, which is lower than the 3.90 times for the model without aggregate shocks. This is because with aggregate shocks, households, especially poor households, have a stronger precautionary saving motive, and their labor income share is thus lower than in the model without aggregate shocks. So the planner would still raise aggregate capital to redistribute through price changes, but not as much as in the economy without aggregate shocks. We provide further validation of this explanation in Section \ref{sec:hh_policy}.

Third, according to Table \ref{table_Davila2} presenting equilibrium statistics conditional on the realization of aggregate shocks, the planner intends for households to increase their savings in a higher ratio (2.84) in the bad aggregate state, compared to (2.74) in the good aggregate state.

\subsection{Impact of Aggregate Shocks on Constrained Optimum}\label{sec:hh_policy}
\begin{figure}[!htb]
     \centering
     \begin{subfigure}[b]{0.54\textwidth}
         \centering
         \includegraphics[width=\textwidth]{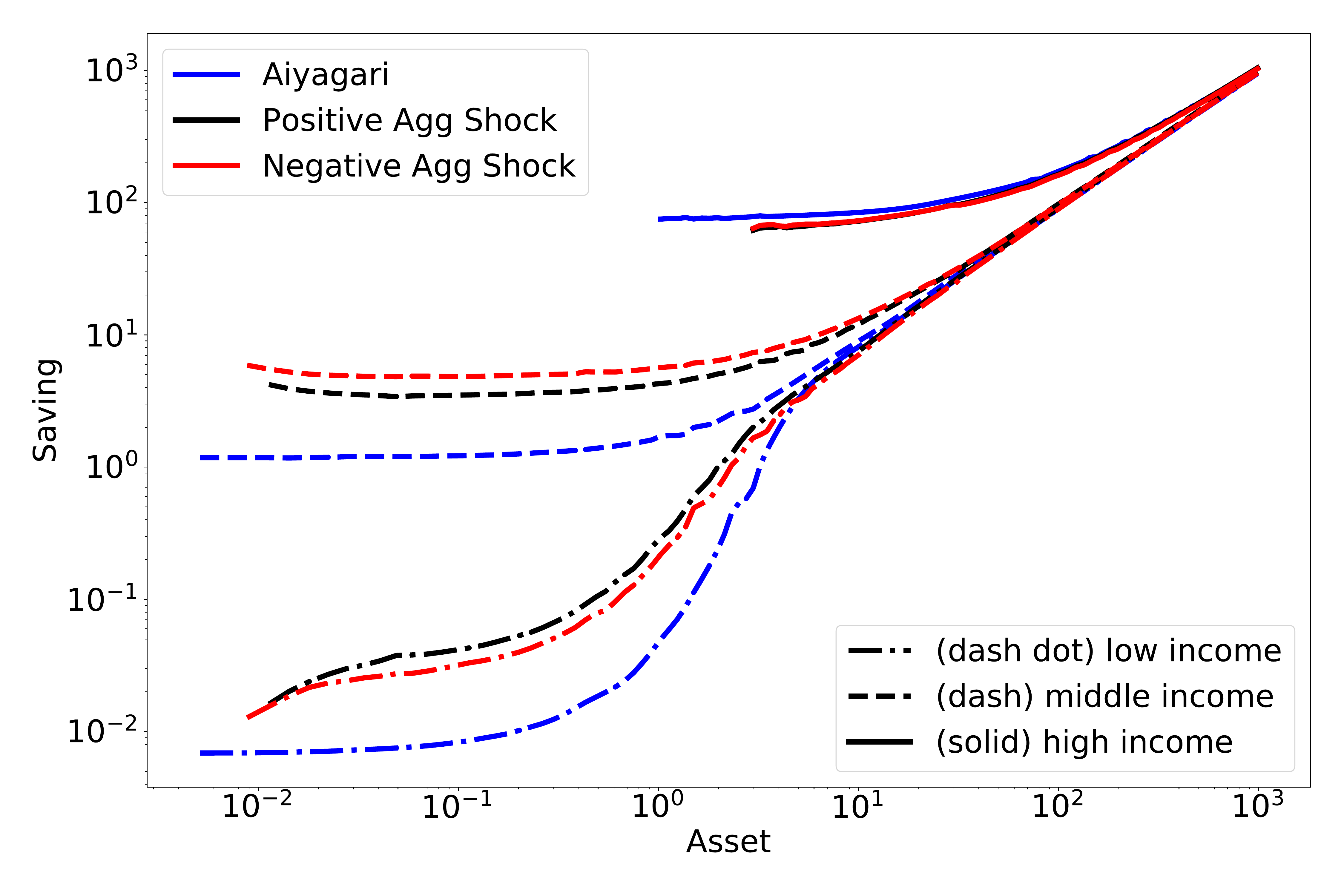}
         \caption{Saving policy}
         \label{fig:saving_policy}
     \end{subfigure}
     \begin{subfigure}[b]{0.45\textwidth}
         \centering
         \includegraphics[width=\textwidth]{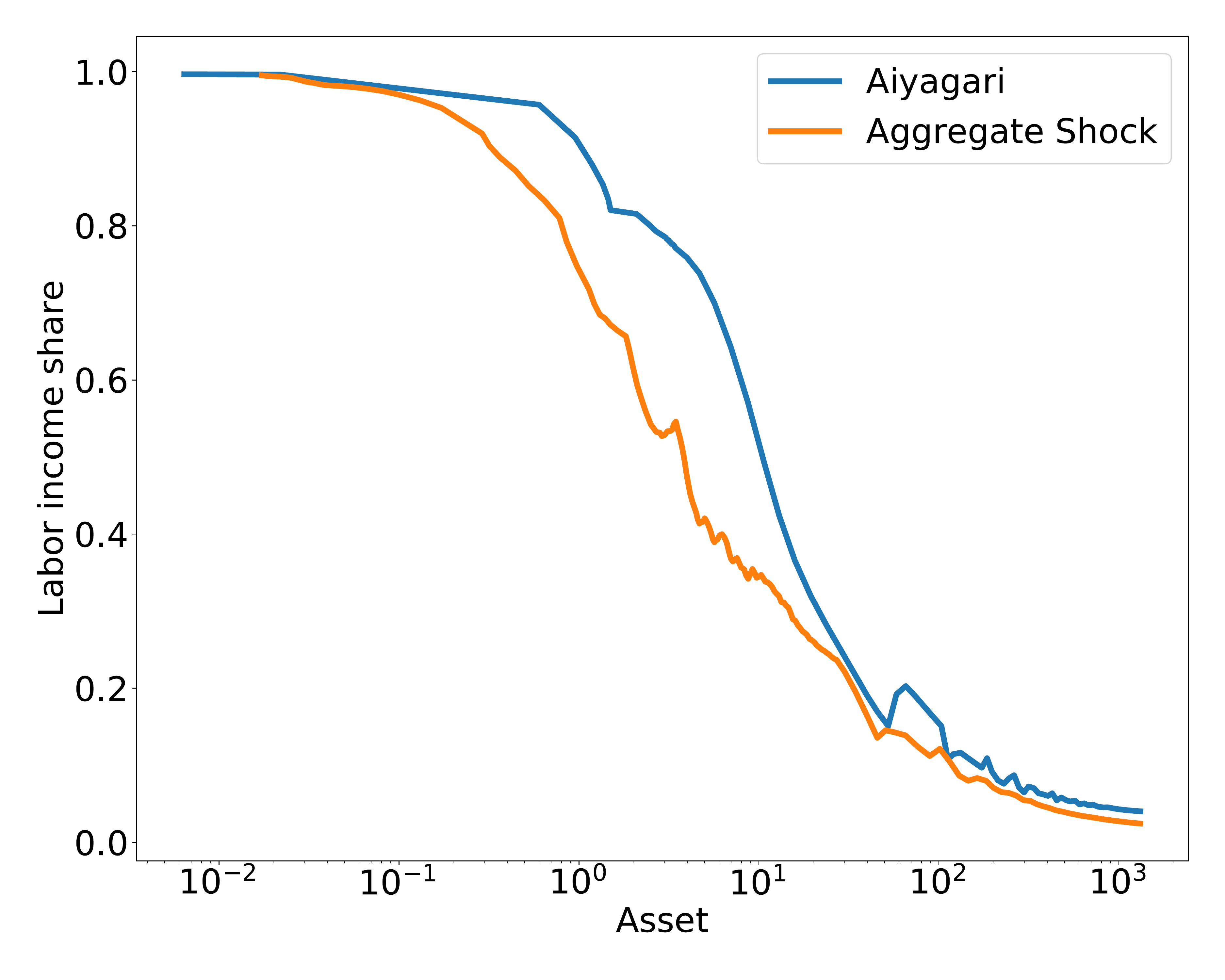}
         \caption{Labor share}
         \label{fig:laborshare}
     \end{subfigure}
  \caption{Household's saving policy and labor share along asset distribution in the constrained optimum. In the model with aggregate shocks, households, especially wealth-poor households, have more precautionary saving and a lower labor income share, compared to the economy without aggregate shocks. }
  \label{fig:policy_laborshare}
\end{figure}

To further understand the impact of aggregate shocks on the constrained efficiency problem, we compare households' saving policy and labor share across the asset distribution in the constrained optimum with aggregate shocks and without aggregate shocks (``Aiyagari economy'') in Figure \ref{fig:policy_laborshare}. Figure \ref{fig:saving_policy} shows that in the presence of aggregate shocks, households, especially wealth-poor households, save more than they do in the Aiyagari economy due to precautionary motives. As a result, households, especially wealth-poor households, have a lower labor income share compared to the Aiyagari economy, which is shown in Figure \ref{fig:laborshare}. As a result, in order to redistribute towards wealth-poor households, the constrained planner does not need to raise the aggregate capital level as much in the presence of aggregate shocks as in the economy without aggregate shocks.

\subsection{Computational Efficiency for Constrained Efficiency Problem}
In this section, we highlight the computational efficiency of DeepHAM in solving the constrained planner's problem. In Table \ref{table:speed_planner}, we report the computational cost for DeepHAM and for the classical method \citep{davila2012constrained} when solving the constrained efficiency problem in the baseline Aiyagari model and in the model with aggregate shocks.

\begin{table}[!htb]  
\centering
\begin{tabular}{ccc}
\hline
\hline
 &  Aiyagari model & With aggregate shocks \\ \hline
Classical method  &
  15 hours &
  not solved in the literature \\ 
DeepHAM &
 20  minutes &
32  minutes \\ \hline\hline
\end{tabular}
\caption{Comparison of the computational speed  for the constrained efficiency problem. The conventional method \citep{davila2012constrained} is implemented on a laptop with a 2.3Ghz Dual-Core Intel Core i5 processor. DeepHAM is implemented on a small cluster with a NVIDIA Tesla P100 GPU.}\label{table:speed_planner}
\end{table}

According to Table \ref{table:speed_planner}, it is very costly to solve the constrained efficiency problem using the classical method, even for the baseline Aiyagari model without aggregate shocks. To our knowledge, a global solution of the constrained efficiency problem in HA models with aggregate shocks has not been presented in the literature. In contrast, DeepHAM can handle this class of problems quite efficiently.\footnote{To be clear, we are not comparing the computational cost of two methods directly since they are implemented on different hardwares. The results mainly show that DeepHAM can be much faster than the conventional method on a modern computing hardware that is easily accessible to researchers.}

\section{Conclusion} \label{sec:conclusion}
In this paper, we present DeepHAM, an efficient, reliable, and interpretable deep learning-based method for globally solving HA models with aggregate shocks. DeepHAM achieves highly accurate results, and can be applied to complex HA models without suffering from the curse of dimensionality. The algorithm automatically generates a flexible and interpretable representation of the agent distribution through generalized moments. The generalized moments extract key information of the distribution that are most relevant to individual decision rules and thus, through aggregation, to welfare and the evolution of the aggregate economy. They furthermore help us better understand whether and how heterogeneity matters in macroeconomics. Moreover, DeepHAM can solve the constrained efficiency problem as fast as solving the competitive equilibrium, a significant advantage over existing methods.
The results demonstrate that DeepHAM is a powerful tool for studying global patterns of complex HA models with aggregate shocks, opening up many exciting possibilities for future research.

In this paper, we assume that the dependence of aggregate prices and quantities on individual states could be written in explicit functional forms. This assumption excludes HA models with aggregate variables that are determined recursively as a function of expected future aggregate variables: the inflation rate in a New Keynesian Phillips curve (NKPC), for example, or indirect utility under Epstein-Zin preferences. Handling forward-looking equilibrium conditions in global solution methods is crucial in solving HANK models with aggregate shocks and requires an additional price function be incorporated in the algorithm. We leave this for further discussion in a companion paper. Another avenue for future research is the development of an estimation algorithm based on DeepHAM.

\singlespacing
\setlength\bibsep{0pt}
\bibliographystyle{te}
\bibliography{HACT}

\onehalfspacing
\appendix
\section*{Appendix}
\section{Neural Networks: a Class of Function Approximator}\label{app:nn}
In this paper, we consider deep, fully connected feedforward neural networks. A network $y = u(x; \Theta)$ with $L ~(L\geq 1)$ hidden layers defines a mapping $\mathbb{R}^{d_1} \rightarrow \mathbb{R}^{d_2}$, in which $x \in \mathbb{R}^{d_1}$ is the input variable, $y \in \mathbb{R}^{d_2}$ is the output variable, and $\Theta=(W_1, b_1, \dots, W_{L+1}, b_{L+1})$ is the collection of network parameters. The network's mapping is defined by a series of compositions of linear transformations and nonlinear activation functions: 
\begin{align*}
    y & = \sigma_{L+1} \circ (W_{L+1} z_L + b_{L+1}), \\
    z_L & =  \sigma_{L-1} \circ (W_{L-1} z_{L-1} + b_{L-1}), \\
    & ~~ \vdots  \\
    z_2 & = \sigma_2 \circ (W_2 z_1 + b_2), \\
    z_1 & = \sigma_1 \circ (W_1 x + b_1).
\end{align*}
Here, $W_l \in \mathbb{R}^{m_{l}\times m_{l-1}}$ is called the weight matrix, and $b_l \in \mathbb{R}^{m_l}$ is called the bias vector, with $l=1,\dots, L+1$. We have $m_0=d_1, m_{L+1}=d_2$, and $m_1, \dots, m_l$ are set as network hyperparameters. $\sigma_l: \mathbb{R} \rightarrow \mathbb{R}$ is a scalar function called an activation function, and $\circ$ denotes element-wise evaluation. The typical choices of $\sigma_l$ include rectified linear units (ReLU) $\sigma(x)=\max\{(0,x)\}$ and the sigmoid function $1/(1+e^{-x})$, among others. Typically, $\sigma_l$ are the same for all $l=1,\dots,L$ and $\sigma_{L+1}$ is chosen as the identity function to ensure the output is unrestricted. In our policy function neural network $c_{\text{NN}}(\cdot)$ in equation \eqref{eq:policy_form}, we choose $\sigma_{L+1}$ as the sigmoid function such that the inequality constraints on the decision variable can be satisfied.

\cite{hornik1989multilayer,cybenko1989approximation} prove that neural networks with one hidden layer neural networks are universal approximators, i.e., they can approximate arbitrary well any unknown Borel measurable function over a compact domain. In recent years, it has been extensively demonstrated empirically and theoretically that deep neural networks with multiple hidden layers have better approximation and optimization efficiency than shallow neural networks with one hidden layer \citep{goodfellow2016deep}. In various fields such as reinforcement learning~\citep{silver2016mastering}, numerical PDEs~\citep{weinan2021algorithms}, and scientific computing~\citep{e2021machine}, deep neural networks have demonstrated astonishing capability in handling high-dimensional state variables in which traditional numerical tools suffer a lot from the curse of dimensionality.

\section{Details of the Model Setup}
\subsection{\cite{krusell1998income} Model}\label{app:KS}

\paragraph{Calibration.} The parameters follow \cite{den2010comparison} with each period representing one quarter. The capital share $\alpha = 0.36$, depreciation rate of capital $\delta = 0.25$, household discount factor $\beta = 0.99$, labor supply $\bar{l} = 1/0.9$, unemployment benefit rate $b = 0.15$, and households have log utility. Aggregate productivity $Z_t \in \{1.01, 0.99\}$. The joint transition matrix for idiosyncratic and aggregate shock is
$\Pi_Z = \begin{bmatrix}
0.525& 0.35& 0.03125& 0.09375\\
0.038889& 0.836111& 0.002083& 0.122917\\
0.09375& 0.03125& 0.291667& 0.583333\\
0.009115& 0.115885& 0.024306& 0.850694
\end{bmatrix}$, where the four rows and columns correspond to $(Z_t, z_t) = (0.99, 0), (0.99, 1), (1.01, 0), (1.01, 1)$ respectively.

\subsection{\cite{fernandez2019financial} Model}\label{app:JFV}

\paragraph{Calibration.} We follow \cite{fernandez2019financial} for parameters. The capital share $\alpha = 0.35$, depreciation rate of capital $\delta = 0.1$, household discount rate $\rho = 0.05$, expert discount rate $\widehat{\rho} = 0.04971$, volatility of aggregate shocks $\sigma = 0.014$, and the risk aversion of the households $\gamma = 2$. To solve the problem in discrete time, we choose $\Delta t = 0.2$, which should be a small number so that the solution is comparable to the continuous time solution. Households' discount factor $\beta = e^{-\rho \Delta t}$. The transition matrix of idiosyncratic shocks is $\Pi_e = \begin{bmatrix}
1 - \lambda_1 \Delta t & \lambda_1 \Delta t \\
\lambda_2 \Delta t & 1 - \lambda_2 \Delta t
\end{bmatrix}$ where $\lambda_1$ = 0.986, $\lambda_1$ = 0.052.

\subsection{\cite{davila2012constrained} Model}\label{app:Davila}
\paragraph{Calibration.} The parameter setting in the baseline model follows \cite{davila2012constrained}.  The capital share $\alpha = 0.36$, depreciation rate of capital $\delta = 0.08$, discount factor $\beta = 0.887$, and the risk aversion of the households $\gamma = 2$. Labor endowment $z^i_{t} \in \{e_0 = 1,  e_1 = 5.29, e_2 = 46.55\}$, and $\Pi_e = \begin{bmatrix}
0.992 & 0.008 & 0\\
    0.009 & 0.980 & 0.011\\
    0 & 0.083 & 0.917
\end{bmatrix}$ with stationary distribution $\{0.498,  0.443, 0.059\}$. The aggregate labor supply $L_t = \bar{L} = 0.498e_0 + 0.443e_1 + 0.059e_3 = 5.574.$\\

In the model with aggregate shocks, $Z_t \in \{Z^l, Z^h\} = \{0.95, 1.05\}$ with transition matrix
$\Pi_Z = \begin{bmatrix}
0.875 & 0.125\\
0.125 & 0.875
\end{bmatrix}$. 
The idiosyncratic shocks are countercyclical, and the transition matrix across labor endowment $\Pi_{e,t} = \begin{bmatrix}
0.98 & 0.02 & 0\\
    0.009 & 0.980 & 0.011\\
    0 & 0.083 & 0.917
\end{bmatrix}$ when $Z_t > 1$, $\Pi_{e,t} = \begin{bmatrix}
0.6512 & 0.3488 & 0\\
    0.978 & 0.011 & 0.011\\
    0 & 0.083 & 0.917
\end{bmatrix}$ when $Z_t < 1$. The aggregate labor supply in good and bad aggregate states are $\{L^h, L^l\} = \{7.525, 3.623\}$, respectively.

\section{Accuracy Measures}\label{app:bellman}
The main accuracy measure we adopt in this paper is the Bellman equation error defined in this section. We choose it over the Euler equation error since it provides a better measure over the whole state space, especially the region close to the inequality constraints, without the need to introduce the Lagrangian multiplier.

In the general HA model in Section \ref{sec:general_setup}, agent $i$'s optimization problem can be characterized recursively:
\begin{equation}\label{bellman1}
V(s^i_{t}, X_t, \bS_t)=\max _{c^i_t}\left[u(c^i_t)+\beta \bbE V(s^i_{t+1}, X_{t+1}, \bS_{t+1})\right].
\end{equation}
Given the solved value function $V(\cdot)$, we can evaluate the Bellman equation error for each state $(s^i_{t}, X_t, \bS_t)$ in the state space as:
\begin{equation}\label{bellman_err1}
\text{err}_{\text{B}}(s^i_{t}, X_t, \bS_t)  = \Big|V(s^i_{t}, X_t, \bS_t) - \max _{c^i_t}\left[u(c^i_t)+\beta \bbE V(s^i_{t+1}, X_{t+1}, \bS_{t+1})\right]\Big|
\end{equation}
where the expectation operator is approximated by Monte Carlo sampling of aggregate and idiosyncratic shocks, and the consumption choice $c^i_t$ is solved again given the solved value function $V(\cdot)$, rather than directly taken from the optimal policy we solve.

We then average with respect to the stationary distribution over $(X_t, \bS_t)$ to calculate the Bellman equation error for the solution we obtain:
\begin{equation}\label{bellman_err}
    \text{err}_{\text{B}} = \bbE_{\mu(\cC^*)}\Big|V(s^i_{t}, X_t, \bS_t) - \max _{c^i_t}\left[u(c^i_t)+\beta \bbE V(s^i_{t+1}, X_{t+1}, \bS_{t+1})\right]\Big|.
\end{equation}

\section{Solution Comparison for Krusell-Smith Model}\label{app:KS_compare}
\begin{figure}[!htb]
\centering
\includegraphics[width=0.49\textwidth]{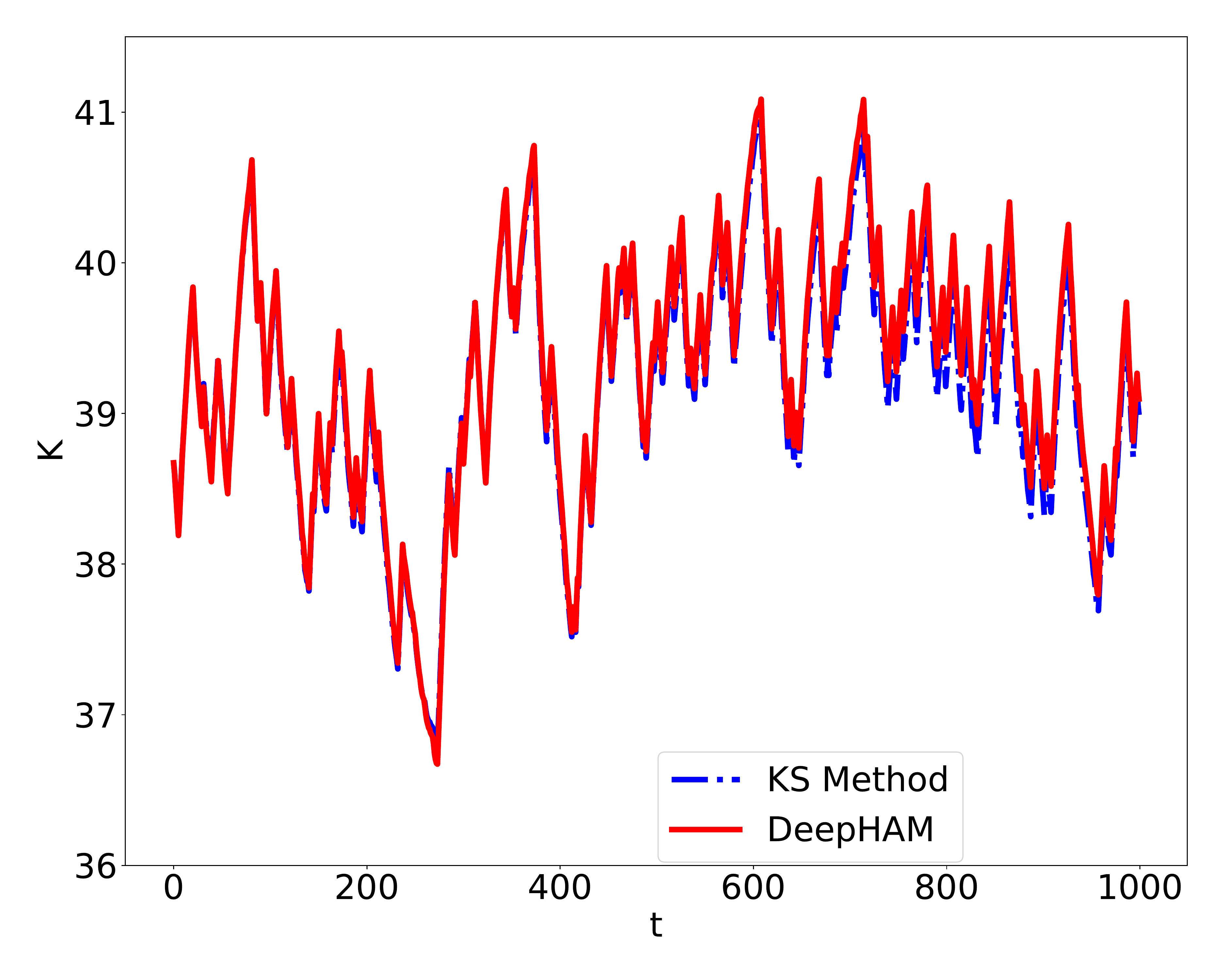}
\includegraphics[width=0.49\textwidth]{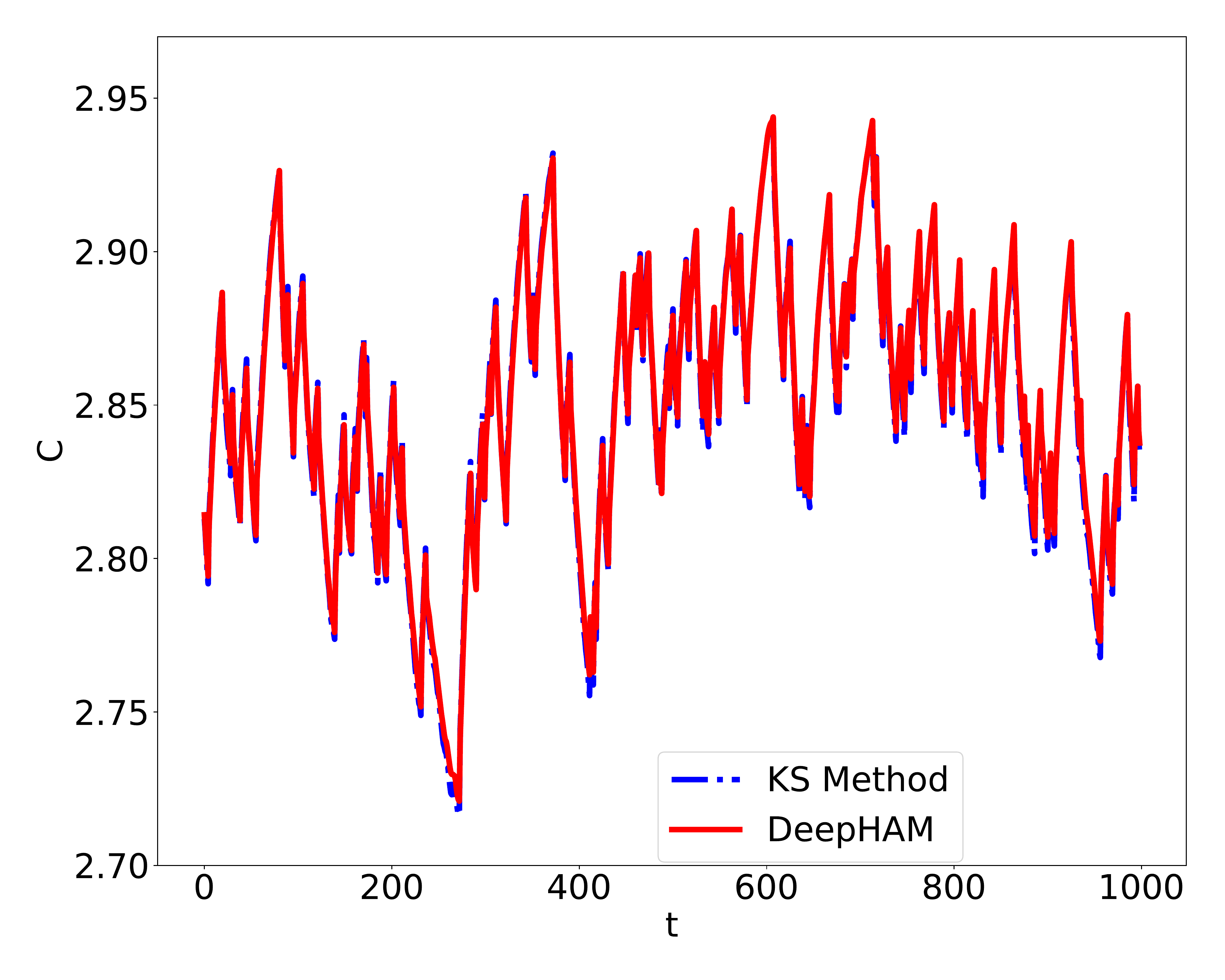}
  \caption{The simulated paths based on solutions obtained from the KS method and DeepHAM, under the same realization of idiosyncratic and aggregate shocks. Left panel: aggregate capital ($K_t$); right panel: aggregate consumption ($C_t$).
  }
  \label{fig:KS_app}
\end{figure}

In this section, we compare the simulated economy based on the DeepHAM solution with the simulation based on the KS method. Under the same realization of idiosyncratic and aggregate shocks, the two economies simulated with 1000 agents based on the two solution methods are presented in Figure~\ref{fig:KS_app}.\footnote{The DeepHAM solution comes from a finite agent model with $N = 50$, while we simulate the two economies with 1000 agents to show the two solutions are consistent with each other in simulations with a relatively large number of agents.}

The KS method with the first moment can solve the Krusell-Smith model reasonably well, as has been widely validated in the literature. Although DeepHAM can further improve the solution accuracy, as we present in Section \ref{sec:KS_acc}, the simulated economy based on the DeepHAM solution is highly consistent with the simulation based on the KS method, as shown in Figure~\ref{fig:KS_app}. This further confirms the accuracy of the DeepHAM solution for the Krusell-Smith model.

\end{document}